\documentclass[prd,aps,
nofootinbib,
floatfix,
superscriptaddress]{revtex4}
\usepackage{graphicx}
\usepackage{epsfig}
\usepackage{rotating}
\usepackage{amssymb}
\usepackage{subfigure}
\usepackage{dsfont}
\usepackage{psfrag}
\usepackage{amsmath,euscript,array,mathrsfs}
\usepackage{axodraw}

\newcommand{\beq}{\begin{equation}}
\newcommand{\eeq}{\end{equation}}
\newcommand{\beqs}{\begin{eqnarray}}
\newcommand{\eeqs}{\end{eqnarray}}
\newcommand{\lsim}{\mathrel{\raisebox{-
.6ex}{$\stackrel{\textstyle<}{\sim}$}}}
\newcommand{\gsim}{\mathrel{\raisebox{-
.6ex}{$\stackrel{\textstyle>}{\sim}$}}}
\newcommand{\Tr}{{\rm Tr}}

\def\hbar{\hspace{0pt}\raisebox{1pt}{$-$} \hspace{-7pt} h}

\def\di{\mbox{d}}

\newcommand{\be}{\begin{equation}}
\newcommand{\ee}{\end{equation}}
\newcommand{\bea}{\begin{eqnarray}}
\newcommand{\eea}{\end{eqnarray}}

\def\lbldef#1#2{\expandafter\gdef\csname #1\endcsname {#2}}

\def\href#1#2{#2}

\newcommand{\ber}{\begin{eqnarray}}
\newcommand{\eer}{\end{eqnarray}}

\newcommand{\beqar}{\begin{eqnarray}}

\newcommand{\eeqar}{\end{eqnarray}}

\newcommand{\dsl}
  {\kern.06em\hbox{\raise.15ex\hbox{$/$}\kern-.56em\hbox{$\partial$}}}

\newcommand{\eeqarr}{\end{eqnarray}}
\newcommand{\ZZ}{{\rm \kern 0.275em Z \kern -0.92em Z}\;}

\def\CC{{\mathchoice
{\rm C\mkern-8mu\vrule height1.45ex depth-.05ex
width.05em\mkern9mu\kern-.05em}
{\rm C\mkern-8mu\vrule height1.45ex depth-.05ex
width.05em\mkern9mu\kern-.05em}
{\rm C\mkern-8mu\vrule height1ex depth-.07ex
width.035em\mkern9mu\kern-.035em}
{\rm C\mkern-8mu\vrule height.65ex depth-.1ex
width.025em\mkern8mu\kern-.025em}}}

\def\RR{{\rm I\kern-1.6pt {\rm R}}}

\def\ZZ{{\rm Z}\kern-3.8pt {\rm Z} \kern2pt}
\def\IB{\relax{\rm I\kern-.18em B}}
\def\ID{\relax{\rm I\kern-.18em D}}
\def\II{\relax{\rm I\kern-.18em I}}
\def\IP{\relax{\rm I\kern-.18em P}}

\newcommand{\bear}{\begin{eqnarray}}
\newcommand{\eear}{\end{eqnarray}}

\def\n{\nu}
  
\def\6{\partial}

\def\bea{\begin{eqnarray}}
\def\eea{\end{eqnarray}}

\def\beqx{\begin{displaymath}}
\def\eeqx{\end{displaymath}}

\newcommand{\bmat}{\left(\begin{array}}
\newcommand{\emat}{\end{array}\right)}

\def\n{\nu}

\def\bo{{\raise-.3ex\hbox{\large$\Box$}}}               
\def\face{{\raise.2ex\hbox{$\displaystyle \bigodot$}\mskip-2.2mu \llap {$\ddot
        \smile$}}}                                   
\def\>{\rangle}                                      
\def\<{\langle}                                      

\def\leftrightarrowfill{$\mathsurround=0pt \mathord\leftarrow \mkern-6mu
        \cleaders\hbox{$\mkern-2mu \mathord- \mkern-2mu$}\hfill
        \mkern-6mu \mathord\rightarrow$}        
\def\dvec#1{\vbox{\ialign{##\crcr
        \leftrightarrowfill\crcr\noalign{\kern-1pt\nointerlineskip}
        $\hfil\displaystyle{#1}\hfil$\crcr}}}           
\def\Tr{{\rm Tr \,}}                                    

\def\-{\hphantom{-}}

\begin{document}
\title{Holographic Technidilaton and LHC searches.
}

\author{Robert Lawrance}
\author{ Maurizio Piai}

\affiliation{Department of Physics, College of Science, Swansea University,
Singleton Park, Swansea, Wales, UK.}

\date{\today}


\begin{abstract}
We analyze in detail the phenomenology of a model of dynamical electroweak symmetry
breaking inspired by walking technicolor, by using the techniques of the bottom-up approach to
holography.
The model admits a light composite scalar state, the dilaton, in the spectrum.
We focus on regions of parameter space for which the mass of such dilaton is 125 GeV,
and for which the bounds on the  precision electroweak parameter S are satisfied.
This  requires that the next-to-lightest composite state is the techni-rho meson,
with a mass larger than 2.3 TeV.
We compute the couplings controlling the decay rates of the dilaton to two photons
and to two (real or virtual) Z and W bosons. For generic choices of the
parameters, we find a  suppression of the decay into heavy gauge bosons,
in respect to the analog decay of the standard-model Higgs.
We find a dramatic effect on the decay into photons, 
which can be both strongly suppressed or strongly enhanced, the latter case corresponding to the 
large-N regime of the dual theory.
There is a correlation between this decay rate of the dilaton into photons 
and the mass splitting between the techni-rho meson and its axial-vector
partner: if the decay is enhanced in respect to the standard-model case, then the 
heavy spin-1 resonances are nearly degenerate in mass, otherwise their 
separation in mass is comparable to the mass scale itself.  
\end{abstract}

\pacs{11.25.Tq, 12.60.Nz.}

\maketitle

\tableofcontents


\section{Introduction}

The main goal of the high-energy physics program of the LHC experiments is to
discover the mechanism responsible for electroweak symmetry breaking 
in the Standard Model of particle physics.
Such mechanism is known to require the existence of new particles and
new interactions, and the LHC has started an extensive program of searches
for  new particles. The LHC can also perform many measurements of 
the properties of the new particles, such as their production and decay rates,
and, by comparing to theoretical predictions, this should allow to reconstruct also the 
new interactions.

One important characterization of new interactions is their strength, which has 
dramatic effects both theoretically and experimentally. While weakly-coupled models 
predict the existence of a small number of new elementary particles, 
with properties that can be computed perturbatively, in strongly-coupled models
one expects a very rich and complicated
 spectrum of bound states, the properties of which are in general very difficult to compute.
However, because of direct and indirect new-physics searches carried out in the last 
thirty years, we know that all the new particles (elementary or not) have to be heavy.
Hence the LHC, particularly in its early stages, will be able to detect at most a handful 
among the lightest such particles.

A particularly nice candidate for a new particle is the Higgs particle in the minimal version of the 
Standard Model. Its properties are very well understood from the theoretical point of view,
with the exception of its mass, which is a free parameter.
The LHC experiments were able, using combinations of data collected during 2011,
to exclude the existence of such particle over most of the allowed parameter
space, with the remarkable exception of a narrow mass range
around $m_s=125$ GeV, where an excess of events over the background
appears to be present in the data~\cite{LHC}.
With more data, it should be possible to establish beyond reasonable doubt whether
a new particle with  mass  $m_s$ is indeed responsible for such excess.
If so, it is also important to measure precisely its properties, in order to establish whether 
it is the Higgs particle of the standard-model  (and hence that the mechanism for 
electroweak symmetry breaking is weakly coupled).
Preliminary studies, which combine all the information collected by the LHC,
are for the time being affected by  large experimental uncertainties, but suggest that some
interesting deviations from the standard-model predictions might be present in the data, for
instance an enhancement of the decay rate into two photons,
and already disfavor some new physics scenarios (for examples of such model-independent analysis 
see~\cite{fits}).

In this paper, we focus our attention on a very special class of strongly-coupled models
of electroweak symmetry breaking (technicolor)~\cite{TC}, which are characterized by the fact that
the lightest composite state is parametrically lighter than the overall strong-coupling scale.
This state is usually called a {\it dilaton} (or techni-dilaton)~\cite{dilaton1,dilaton2,dilaton3,dilatonpheno,dilaton4,dilaton5D,dilatonnew}, to highlight  the fact  that the reason why 
such scalar is light is related to the spontaneous breaking of dilatation invariance.
The fundamental theory is a strongly-coupled gauge theory with new matter fields (techni-quarks),
such that just above the confinement scale the dynamics is approximately conformal, and the running
of the couplings is slow (walking)~\cite{WTC,reviews}.
Besides the fact that the approximate scale invariance produces a light scalar in the spectrum, 
these models are particularly appealing because of their multi-scale nature, 
and of the fact that large anomalous dimensions arise in the study of the renormalization group flow,
all of which allows to address generic problems of strongly-coupled models with
precision electro-weak physics, with flavor physics and with the generation of fermion masses.

What makes the (techni-)dilaton particularly intriguing is that 
its properties are  similar to those of the Higgs particle of the Standard Model.
Hence, if the signal at 125 GeV is confirmed, the new particle might  be 
one such techni-dilaton, in which case electroweak symmetry breaking would arise from 
a strongly-coupled model, with properties diametrically opposite to the ones of
the minimal version of the Standard Model.
In order to disentangle this ambiguity, a precise set of measurements of the 
properties of the new particle is needed, together with equally precise theoretical
calculations, in particular of all its production and decay rates.

Because of the strong coupling, performing such calculations is highly non-trivial.
A helpful tool in this direction is provided by the ideas of gauge-gravity dualities~\cite{AdSCFT,reviewAdSCFT}.
The basic concept is that strongly-coupled field theories can be reformulated
in terms of  dual weakly-coupled gravity theories in higher space-time dimensions.
Because the dual theories are weakly-coupled, the calculability problems are
overcome, and hence one could compute precisely the properties of the techni-dilaton,
once the gravity dual of a walking technicolor theory is known.
Still, it is not easy to find the exact dual of a given gauge theory, particularly when 
the field theory confines, has a non-trivial multi-scale dynamics and is not supersymmetric
(for recent encouraging results along this line, see for instance the family 
of models in~\cite{stringWTC}).

In this paper, we take a more pragmatic approach, along the lines 
of earlier studies on holographic technicolor within the bottom-up approach~\cite{AdSTC}.
We assume that a walking technicolor theory exists,
without specifying its elementary properties.
We assume also that it admits a gravity dual, 
but rather than specifying its complete 10-dimensional string theory
formulation, we restrict attention to its truncated 5-dimensional gravity description,
which we choose on the basis of very generic and simple properties.
Namely, we know that such theory must admit a region with walking dynamics,
it must confine, and it must induce electroweak symmetry breaking.
We focus on the simplest possible model, based on a five-dimensional sigma-model
scalar field coupled to gravity, which allows to reproduce these generic 
requirements~\cite{EPdilaton}.
We then perform a detailed analysis of the phenomenology of the model which results,
as a function of its free parameters.

As we will see, in spite of its simplicity, the model that results naturally satisfies
all the direct and indirect bounds on electroweak physics,
in particular those from the $S$ parameter~\cite{Peskin,Barbieri},
 and predicts the existence of a light dilaton,
the mass of which can be chosen to agree with $m_s=125$ GeV.
The phenomenological properties of such scalar are similar to those of the Higgs 
particle of the minimal version of the Standard Model.  However, 
for generic and natural choices of the parameters, the model allows to make 
predictions for many other LHC observables, such that it can be distinguished, 
on the basis of experimental data
to be collected by the LHC, from the minimal version of the standard model.

In reference to the predictions for the observable quantities, our results should be interpreted in
a spirit similar to that of a generic low energy effective theory originating from a class of unknown theories.
Our results will tell us what is possible and what is not, on general grounds.
But they will also tell us what is a generic expectation, true over large 
parts of the parameter space (and hence for a large number of possible microscopic realizations
of this scenario), and what happens only for contrived choices of the parameters (and hence only
for very special, possibly unnatural microscopic theories).
Finally, again in the spirit similar to the one of effective theories,  
we will also discover the existence of non-trivial correlations between superficially
unrelated observables.

With more data, the LHC will be able to tell whether this class of models is favored or disfavored
in respect to the minimal version of the Standard Model. In the former case, it would become
an immediate priority for the community to look for fundamental theories that are fully calculable and
successfully reproduce the low-energy results, both in field theory and within the rigorous construction
of their string-theory dual description.

\subsection{The Higgs particle in the Standard Model.}

In this brief introductory subsection, we recall some useful results about the minimal standard-model 
Higgs and its decays.
First of all, the Higgs field in the minimal Standard Model is a scalar $H$ transforming as 
$(1,2,1/2)$ under the $SU(3)_c\times SU(2)_L\times U(1)_Y$ gauge group.
Symmetry breaking to $SU(3)_c\times U(1)_{e.m.}$ is triggered by assuming a potential for the Higgs field
\beqs
V&=&\lambda\left(H^{\dagger}H-\frac{v_W^2}{{2}}\right)^2\,.
\eeqs
As long as $\lambda>0$, the minimum of the potential yields the Vacuum Expectation Value (VEV)
  $\langle H^{\dagger}H\rangle = \frac{v_{W}^2}{2}$,
and hence in unitary gauge we have
\beqs
H&=&\frac{v_W+h}{\sqrt{2}}\left(\begin{array}{c} 0\cr 1\end{array}\right)\,,
\eeqs
with $h$ the canonically normalized field describing the massive scalar fluctuations around the minimum.
The Higgs particle  has a mass $m_h^2=2\lambda v_W^2$.

In this expression, we notice a few important things. 
First of all the mass of the Higgs
depends on the scale fixed by the VEV, which can be rewritten in terms of the Fermi decay constant
 $\frac{G_F}{\sqrt{2}}=\frac{1}{2v_{W}^2}$.
 As long as $\lambda$ is small, the classical Lagrangian is approximately scale invariant.
 Scale invariance is broken spontaneously by the VEV $v_W$, and explicitly by $\lambda$,
 and hence the Higgs particle can be thought of as a pseudo-dilaton, the presence of which arises from
 the spontaneous breaking of scale invariance at the electroweak scale, and 
 the mass of which is controlled by explicit symmetry breaking, which is parameterized by $\lambda$.
 The smallness of $\lambda$ has hence two technical consequences: 
 it ensures that perturbation theory be viable, but it also makes the effects of
  the explicit breaking of scale invariance small.
  
  At the tree-level, by expanding the Lagrangian at the linear order in $h$ we find the following couplings
  \beqs
  {\cal L}&=&2\frac{h}{v_W}M_W^2W^{+}_{\mu}W^{-\,\mu}\,+\,\frac{h}{v_W}M_Z^2 Z_{\mu}Z^{\mu}
  -\frac{h}{v_W}M_{\psi}\bar{\psi}\psi\,+\,\cdots\,,
  \eeqs
  where $W$ and $Z$ are the massive gauge bosons of the standard model, and $\psi$
  the massive fermions.
  Notice how all these couplings are inversely proportional to the decay constant $v_W$,
  and proportional to the masses of the particles $h$ interacts with, which is a consequence
  of the pseudo-dilaton nature of $h$.
  These couplings control the main decay channels contributing to the Higgs width.
  
 At the quantum level, there is another source of explicit breaking of scale invariance: the fact that
 the coupling constants actually run, because all the operators in the quantum theory are quasi-marginal.
 This is in general a very small effect, which adds only tiny corrections to most of the couplings.
 But it becomes important for the photon and gluons, which do not couple
 directly at the tree-level being massless.
 One important consequence of this is that the neutral Higgs particle decays into two-photons,
with a coupling strength that is related to the contribution of the heavy particles to 
the conformal anomaly of QED. 
 The dominant contributions to such decay come from 1-loop diagrams
 involving charged heavy particles, which in
  the Standard Model are the $W$ boson and the top quark.
 The resulting  decay rate is:
 \beqs
 \Gamma(h\rightarrow \gamma\gamma)_{SM}&=&\frac{G_F\alpha^2m_h^3}{128\sqrt{2}\pi^3}
 \left|\sum_fN_cQ_f^2 A_{f}(\tau_f)+A_{W}(\tau_W)\frac{}{}\right|^2\,,
 \eeqs
 where $\alpha$ is the QED coupling, $N_c=3$ and $Q_f$ is the charge of the fermion of species $f$ (the top quark), while $\tau_i=m_h^2/(4M_i^2)$.
 The functions $A_f$ and $A_W$, for $\tau\leq 1$, are given by~\cite{reviewHiggs}
 \beqs
 A_f(\tau)&=&\frac{2}{\tau^2}\left(\tau+(\tau-1)\arcsin^2\sqrt{\tau}\right)\,,\\
 A_1(\tau)&=&-\frac{1}{\tau^2}\left(2\tau^2+3\tau+3(2\tau-1)\arcsin^2\sqrt{\tau}\right)\,.
 \eeqs
 Notice that the contribution 
 of a new heavy fermion to the amplitude is controlled by
 \beqs
 \lim_{\tau\rightarrow 0} A_f(\tau)&=&\frac{4}{3}\,.
 \eeqs
 
All of this goes to show one very special thing 
 about the minimal version of the Standard Model: the only free parameter 
 is the mass of the Higgs $m_h$
 (or equivalently $\lambda$) and hence once the mass is measured, 
 all the decay rates and production rates are determined, up the potential difficulties 
 involved with the fact that the initial state at the LHC consists of protons, 
 rather than the partons the couplings of which are known.
 
 Finally, let us remind the reader of the result of a 1-loop perturbative 
 exercise. Consider  the effect on the running of the electromagnetic 
 gauge coupling due to
 a set $n_f$ of fermions of change $Q_f$. This yields
 \beqs
 \alpha(\mu)&=&\frac{\alpha}{1-\frac{\alpha}{3\pi}n_f Q^2\ln(\mu^2/M^2)}\,,
 \eeqs
 where $M$ is a reference scale chosen so that $\alpha(M)=\alpha$
  and $\mu$ is the renormalization scale.
 For later convenience, we {\it define}
 \beqs
 \beta&\equiv&\frac{2\alpha}{3\pi}n_fQ^2\,.
 \eeqs
 Let us now assume that all of these fermions are heavier than the Higgs particle,
  so that $\tau_f\ll 1$. This is a very good
 approximation even for the top quark (for which $n_f=N_c=3$ and $Q=2/3$),
 and  we can rewrite
 \beqs
  \Gamma(h\rightarrow \gamma\gamma)_{SM}&=&\frac{G_F\alpha^2m_h^3}{128\sqrt{2}\pi^3}
 \left|2\pi \frac{\beta}{\alpha}+A_{W}(\tau_W)\frac{}{}\right|^2\,,
 \eeqs
 where $A_W\simeq -8.3$ by using the experimental value of the $W$-boson mass
 (notice that $M_W$ is not very large, $\tau_W\simeq {\cal O}(1)$ for $m_h\simeq 125$ GeV).
 
  As a final comment, the $W$ and top contributions 
  enter with opposite signs into the decay amplitude. In particular, if there were a few new 
 heavy fermions besides the top,  
 they would contribute to the fermionic part of the amplitude, and suppress the decay.
For example, the addition of a whole family of new fermions with quantum numbers 
replicating those of the standard-model ordinary quarks and leptons would result
in a strong suppression of the decay rate, which would render the 
detection of this decay very difficult for the LHC. 

\section{The Model}

\subsection{Geometry}

Using the ideas of holography, we build a five-dimensional model, dual to a four-dimensional theory of walking technicolor (WTC). The metric of the five-dimensional space-time is, in full generality, given by
\begin{equation}\label{metric}
ds^2=e^{2A(r)}\eta_{\mu\nu}dx^\mu\,dx^\nu+dr^2,
\end{equation}
where $r$ is the extra dimension, $A(r)$ is the warp factor, capital Roman indices span 
$M,N=0,1,2,3,4$, lower case Greek indices span $\mu,\nu=0,1,2,3$ and $\eta_{\mu\nu}$ has signature (-,+,+,+). 
$A(r)$ is independent of the space-time directions $x^\mu$ and, if $A(r)$ is linear, we recover 
an AdS space. To this we add two $3+1$ dimensional boundaries, an IR boundary at $r=r_1$ which mimics
 the confinement scale of the theory and a UV boundary at $r=r_2$. 
 The UV boundary acts as a regulator and the limit $r_2\rightarrow\infty$ 
 should be taken in subsequent calculations. 

\subsection{Bulk Scalars and the Classical Background}

The matter content of this model consists of a single bulk sigma-model scalar and a set of bulk $SU(2)_L\times SU(2)_R$ gauge bosons.~\footnote{Aside from adding by hand the contribution from top loops
to the coupling of the dilaton to photons and gluons,  in this paper we ignore the SM fermions altogether.}
 The action for the bulk scalar coupled to gravity is~\cite{EP}
\begin{eqnarray}
\mathcal{S}&=&\int d^4x dr \sqrt{-g}\Theta\left(\frac{R}{4}+\mathcal{L}_5\right)+\sqrt{-\tilde{g}}\delta(r-r_1)\left(\frac{K}{2}+\mathcal{L}_1\right)\nonumber\\&&-\sqrt{-\tilde{g}}\delta(r-r_2)\left(\frac{K}{2}+\mathcal{L}_2\right).
\end{eqnarray}
where $R$ is the Ricci scalar and $K$ is the extrinsic curvature of the boundary hyper-surface, defined by
\begin{eqnarray}
K_{MN}&=&\nabla_M N_N\,,\,\,\,\,
K\,=\,g^{MN}K_{MN}.
\end{eqnarray}
$N_N$ is an orthonormal vector to the surface, and
\begin{eqnarray}
\mathcal{L}_5&=&-\frac{1}{2}g^{MN}\partial_M\Phi\partial_N\Phi-V(\Phi)\,,\,\,\,\,
\mathcal{L}_1\,=\,-\lambda_1(\Phi)\,,\,\,\,\,
\mathcal{L}_2\,=\,-\lambda_2(\Phi)\,,
\end{eqnarray}
where $V(\Phi)$ is a bulk potential and the $\lambda_i(\Phi)$ are localized potentials, living on the 4D boundaries. By varying the bulk part of this action with respect to the metric the Einstein equations of the system can be derived, yielding
\begin{eqnarray}\label{eqn::scalarEE}
3A^{\prime\prime}+6(A^\prime)^2&=&-\bar{\Phi}^{\prime 2}-2V,\\
6(A^\prime)^2&=&\bar{\Phi}^{\prime 2}-2V,
\end{eqnarray}
where the bar denotes the classical solution and primes denote differentiation with respect to the extra dimension $r$. Whilst varying with respect to the field $\Phi$ gives the equations of motion
\begin{equation}\label{eqn:scalarEL}
\bar{\Phi}^{\prime\prime}+4A^\prime\bar{\Phi}^{\prime}+\bar{\Phi}^{\prime 2}-\partial_\Phi V=0.
\end{equation}
Boundary conditions for the background solutions can also be found by considering the variation of the full action. This yields
\begin{eqnarray}
\bar{\Phi}^{\prime}|_{r_i}&=&\partial_\Phi\lambda_i|_{r_i}\\
A^\prime|_{r_i}&=&-\frac{2}{3}\lambda_i|_{r_i},
\end{eqnarray}
and if the potential $V$ is such that it can be written in terms of a superpotential,
\begin{equation}
V=\frac{1}{2}(\partial_\Phi W)^2-\frac{4}{3}W^2.
\end{equation}
The $\lambda_i$ can be expanded in terms of this superpotential
\begin{equation}
\lambda_i=W(\Phi_i)+\partial_\Phi W|_{\Phi_i}(\Phi-\Phi_i)+....
\end{equation}
and we find
\begin{equation}\label{eqn:Aprime}
A^\prime=-\frac{2}{3}W,
\end{equation}
and
\begin{equation}\label{Phiprime}
\bar{\Phi}^{\prime}=\partial_\Phi W.
\end{equation}

\begin{figure}[t]
\begin{center}
\subfigure{\includegraphics[scale=0.7]{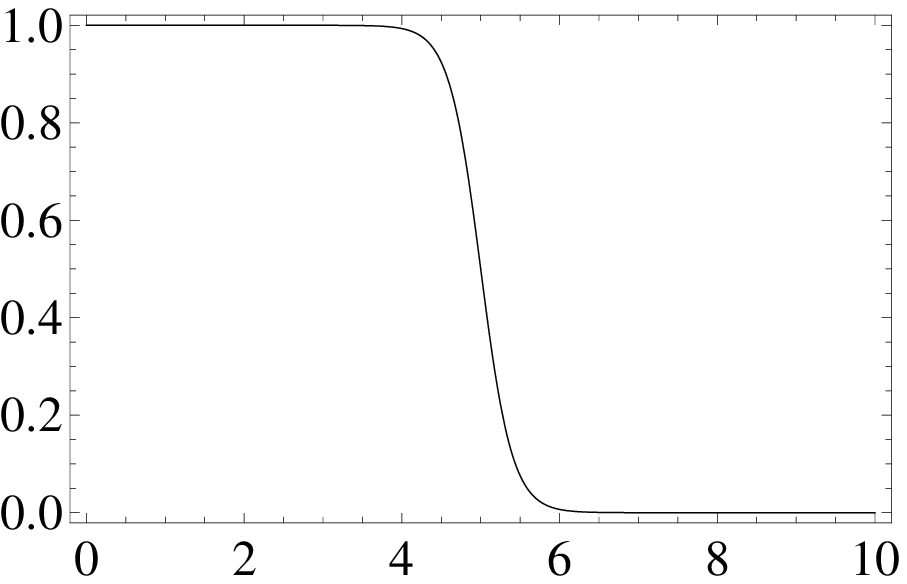}
\Text(-190,110)[]{\small{$\bar{\Phi}$}}
\Text(-20,0)[]{\small{$r$}}}
\subfigure{\includegraphics[scale=0.7]{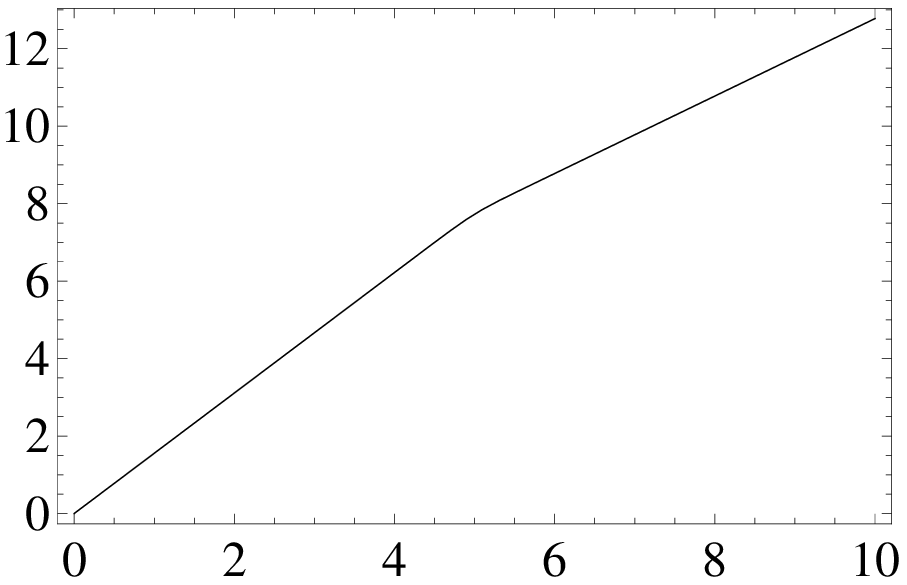}
\Text(10,110)[]{\small{$A$}}
\Text(-20,0)[]{\small{$r$}}}
\end{center}\caption{Left panel: plot of $\bar{\Phi}$ against $r$ for $\Phi_I=1$, $\Delta=5$ and $r_*=5$. Note that $\bar{\Phi}$ is approximately constant in the two regions $r<r_*$ and $r>r_*$. Right panel: plot of $A(r)$ against $r$ for $\Phi_I=1$, $\Delta=5$ and $r_*=5$. Note that $A(r)$ approximately linear in the two regions $r<r_*$ and $r>r_*$.}
\label{phibarAplot}
\end{figure}

The superpotential we consider is cubic and has the form~\cite{EPdilaton}
\begin{equation}\label{superpotential}
W(\Phi)=-\frac{3}{2}-\frac{\Delta}{2}\Phi^2+\frac{\Delta}{3\Phi_I}\Phi^3,
\end{equation}
which means that there are two stable critical points at $\Phi=0$ and $\Phi=\Phi_I$,
i.~e. this system represents the dual of a theory which admits to different fixed points.
For simplicity, in the following we will often focus on solutions for the choice $\Delta=\Phi_I=1$. 
Solving Eq.~(\ref{Phiprime}) for this choice of superpotential gives
\begin{equation}
\bar{\Phi}=\frac{\Phi_I}{1+e^{\Delta(r-r_*)}},
\end{equation}
which is shown in the left panel of Figure \ref{phibarAplot}. Note that $\bar{\Phi}$ is approximately constant in the regions $r<r_*$ and $r>r_*$. This also means that $W$ and its derivatives with respect to $\Phi$ are also approximately constant in these two regions. $A(r)$ can be found exactly from Eq.~(\ref{eqn:Aprime}) and the right panel of Figure \ref{phibarAplot} shows this exact solution. We will also find it useful to approximate $A(r)$ as
\begin{equation}\label{aprox}
A=\left\{\begin{array}{l}-\frac{2}{3}W_1r,\,r<r_*\\-\frac{2}{3}W_2(r-(1-\frac{W_1}{W_2})r_*),\,r>r_*\end{array}\right.,
\end{equation}
where $W_1$ is the value of $W$ in the region $r<r_*$ and $W_2$ the value in the region $r>r_*$. A warp factor of this form describes a space which is approximately AdS in each region, but where the curvature of the space goes through a transition at the point $r=r_*$.
Approximately, we have
\beqs
W_1&=&-\frac{3}{2}\left(1+\frac{\Delta\Phi_I^2}{9}\right)\,\,,\,W_2\,=\,-\frac{3}{2}\,.
\eeqs

\subsection{Gauge Sector}

The gauge sector we consider is $SU(2)_L\times SU(2)_R$ and chiral symmetry breaking 
$SU(2)_L\times SU(2)_R\rightarrow SU(2)_V$ occurs due to different  IR 
boundary conditions for vector and axial-vector fields. We introduce the gauge bosons in the probe approximation, so that they do not backreact on the metric. They are introduced via the action
\begin{equation}\label{eqn:gaugeaction}
\mathcal{S}_{gauge}=-\frac{1}{4}\int d^4x\int^{r_2}_{r_1}dr\, \left(\frac{}{}a(r)-D b(r)\delta(r-r_2)\right)F^a_{\mu\nu}F_a^{\mu\nu}+2b(r)F^a_{r\mu}F_a^{r\mu}-2b(r){\Omega^2}{}\mathcal{W}^{a\mu} \mathcal{W}^a_\mu\delta(r-r_1)\,,
\end{equation}
where we have included a VEV $\Omega$ in the IR and a UV-boundary kinetic term $D$,
and we represent both left and right handed groups by the field-strength tensors. For $SU(2)$ we have that the (non-abelian) field strength tensor is given by
\begin{equation}
F^a_{MN}=\partial_M \mathcal{W}^a_N-\partial_N \mathcal{W}^a_M+ig\varepsilon^{abc}\mathcal{W}^b_M \mathcal{W}^c_N\,,
\end{equation}
where  $g$ is the dimensionful gauge coupling. 
The functions $a(r)$ and $b(r)$ enter the action due to the fact that the space-time the gauge bosons propagate in is curved. This means that sums over space-time indices should be modified by the appropriate metric elements and a $\sqrt{-g}$ term. $a(r)$ and $b(r)$  parametrise these modifications and have the form
\begin{eqnarray}
a(r)&=&1,\\
b(r)&=&e^{2A}.
\end{eqnarray}
The equations of motion and boundary conditions can now be derived using the variational principle. 


After Fourier transforming the 4D space-time directions and applying the (unitary) gauge choice $\mathcal{W}_r=0$, it can be shown that the transverse components satisfy
 the bulk equation~\footnote{Here and in the following we  define the 
 four-momentum $q^2=-\eta_{\mu\n}q^{\mu}q^{\nu}$.
This seemingly bizarre choice, and a related redefinition of the signs in the vacuum polarization tensors
which we discuss later in the paper,
have the function of facilitating the comparison with the literature. When discussing 
string-theory and supergravity sigma-models, it is customary to adopt the convention in which
the metric has signature $\{-,+,+,+\}$, as we did here. However, when discussing the phenomenology
of Kaluza-Klein theories and of precision electroweak measurements, it is customary to adopt the
convention where the metric has signature $\{+,-,-,-\}$. With this change of sign in $q^2$,
all the equations agree with those in the literature. In particular, the on-shell condition for a particle of mass
$m$ is $q^2=m^2>0$.}
\begin{equation}
a(r)q^2\mathcal{W}_\nu^a(q^2,r)+\partial_r(b(r)\partial^r\mathcal{W}_\nu^a(q^2,r))=0,
\end{equation}
while the IR boundary conditions are
\begin{eqnarray}
\partial_r\mathcal{W}_\nu^a(q^2,r)|_{r_1}&=&0\,,\\
(\partial^r\mathcal{W}_\nu^a(q^2,r)-\Omega^2\mathcal{W}_\nu^a(q^2,r))|_{r_1}&=&0.
\end{eqnarray}
Working in the vector/axial basis and following the notation of~\cite{EPdilaton}, one can then write
\begin{equation}
\mathcal{W}_\nu^a(q^2,r)=v^a(q^2,r)\mathcal{W}_\nu^a(q^2),
\end{equation}
and define $\partial_r v^a(q^2,r)\equiv\gamma^a(q^2,r)v^a(q^2,r)$ for the vectors and $v^a(q^2,r)\equiv\chi^a (q^2,r)\partial_r v^a(q^2,r)$ for the axial-vectors, so that the bulk equation and boundary conditions become
\begin{eqnarray}
\partial_r(b(r)\gamma^a(q^2,r))+b(r)(\gamma^a(q^2,r))^2+a(r)q^2&=&0,\label{eqn:gammabulk}\\
-b^2(r)\partial_r\left(\frac{\chi^a(q^2,r)}{b(r)}\right)+b(r)+a(r)q^2(\chi^a(q^2,r))^2&=&0,
\end{eqnarray}
and
\begin{eqnarray}
\gamma^a(q^2,r_1)&=&0,\label{s1bcs}\\
1-\Omega^2\chi^a(q^2,r_1)&=&0,
\end{eqnarray}
respectively. Note that, when $\Omega$ is different from 0, vector and axial-vector solutions have different IR boundary conditions, which triggers EWSB. 
Whilst, for $\Omega=0$, the second of these equations can 
only be satisfied for $\chi^a(q^2,r_1)=\infty$. Since $\gamma$ and $\chi$ are related by
$
\gamma^a(q^2,r)=(\chi^a(q^2,r))^{-1},
$
this is the case where electroweak symmetry remains unbroken.

\section{Precision Physics}

\subsection{Holographic Renormalization and Counter-terms}

To analyse the spin-1 sector it is first necessary to find the (holographically renormalized~\cite{HR})
vacuum polarization tensors
 defined by the effective 4D action
\begin{equation}\label{4deffs}
\mathcal{S}_4=\int d^4x\,\left\{\frac{1}{2}P^{\mu\nu}\mathcal{W}_\mu^i(-q)\pi_{ij}(q^2)\mathcal{W}_\nu^j(q)\right\}\,,
\end{equation}
with $P^{\mu\nu}=\eta_{\mu\nu}+q^{\mu}q^{\nu}/q^2$.~\footnote{Notice again two changes of sign
here. First of all, the appearance of a $+$ in the relative sign between the 
two terms in $P^{\mu\nu}$ is due to the change of sign in $q^2$ and ensures that $P^{\mu\nu}q_{\mu}=0$.
But also, we changed the overall sign of the right-hand side of Eq.~(\ref{4deffs}), in such a way that
the functions $\pi$ are exactly those that would appear by adopting the convention where the signature
is $\{+,-,-,-\}$ in the metric.}
In the vector-axial basis the matrix $\pi_{ij}(q^2)$ has the form
\begin{equation}
\pi_{ij}(q^2)=\left(\begin{array}{cc}\pi_A & 0 \\0 & \pi_V\end{array}\right),
\end{equation}
and  it can be shown that
\begin{eqnarray}
\pi_A&=&-\mathcal{N}(r_2)\left(q^2Db(r_2)+\frac{b(r_2)}{\chi(q^2,r_2)}\right),\nonumber\\
\pi_V&=&-\mathcal{N}(r_2)\left(q^2Db(r_2)+b(r_2)\gamma(q^2,r_2)\right),
\end{eqnarray}
where $Db(r_2)$ is a counter-term. To find the form of this counter-term, first we expand $\pi_V$ in $q^2$ as
\begin{equation}\label{exppi}
\pi_V(q^2)=\pi_V(0)+q^2\pi^\prime_V(0)+\frac{1}{2}q^4\pi^{\prime\prime}_V(0)+...,
\end{equation}
and the function $\gamma(q^2,r)$ as
\begin{equation}\label{expg}
\gamma(q^2,r)=\gamma_0(r)+q^2\gamma_1(r)+\frac{1}{2}q^4\gamma_2(r)+....
\end{equation}
This then implies
\begin{equation}
\pi^{\prime}_V=-\mathcal{N}(r_2)(Db(r_2)+b(r_2)\gamma_1(r_2)),
\end{equation}
and we can find $\gamma_1$ by expanding $\gamma$ in Eq.~(\ref{eqn:gammabulk}), giving
\begin{eqnarray}
\partial_r(b(r)\gamma_0)+b(r)\gamma_0^2&=&0,\\
\partial_r(b(r)\gamma_1)+2b(r)\gamma_0\gamma_1+a(r)&=&0.
\end{eqnarray}
The first of these equations, in combination with the IR boundary condition Eq.~(\ref{s1bcs}), 
implies that $\gamma_0=0$ for all $r$, which indicates the presence of a 
massless vector state --- the photon. We also find that
\begin{equation}
\gamma_1(r_2)=-\frac{1}{b(r_2)}\int\limits_{r_1}^{r_2}dr^\prime a(r^\prime)=-\frac{r_2}{b(r_2)},
\end{equation}
which indicates the presence of a divergence in $\pi_V$ for $r_2\rightarrow\infty$. 
Choosing the counter-term to be
\begin{equation}\label{CT}
Db(r_2)=r_2-\frac{1}{\varepsilon^2},
\end{equation}
where $\varepsilon$ is a (finite) free parameter, 
 the divergence cancels. 
 Selecting the normalization $\pi_V^\prime(0)=1$ determines 
 $\mathcal{N}(r_2)=\varepsilon^2$, so that the vacuum polarizations are finally given by
\begin{equation}\label{corv}
\pi_A(q^2)=-\varepsilon^2\left(q^2\left(r_2-\frac{1}{\varepsilon^2}\right)+\frac{b(r_2)}{\chi(q^2,r_2)}\right)\,,
\end{equation}
and
\begin{equation}\label{cora}
\pi_V(q^2)=-\varepsilon^2\left(q^2\left(r_2-\frac{1}{\varepsilon^2}\right)+b(r_2)\gamma(q^2,r_2)\right)\,,
\end{equation}
where it is understood that we have to take the limit $r_2\rightarrow\infty$. 

\subsection{Electroweak Precision Parameters}

The electroweak precision parameters (EWPT) $\hat{S},\hat{T},U,X$ and $Y$ \cite{Barbieri} parametrise the experimental bounds on new physics contributions to the standard model. 
For small values of $\varepsilon$ we need only to concentrate on $\hat{S}$ and $\hat{T}$, 
since satisfying these will ensure that the others are also satisfied~\cite{AdSTC}. 
$\hat{S}$ and $\hat{T}$ are given by
\begin{equation}
\hat{S}\,\equiv\,\cos^2\theta_W(\pi^\prime_{V^3}(0)-\pi^\prime_{A^3(0)})\,\lsim\,3\times 10^{-3},
\end{equation}
and
\begin{equation}
\hat{T}\,\equiv\,\frac{1}{M_W^2}(\pi_{33}(0)-\pi_{+-}(0))\,\lsim\,5\times 10^{-3},
\end{equation}
respectively. The bounds are indicative $3\sigma$-level bounds obtained by extrapolating
 over the allowed range of the SM mass of the Higgs $m_h$.

The parameter $\hat{T}$  measures the splitting between the squared masses of the $W^3$ and $W^\pm$. 
 Because of the gauging of the whole $SU(2)\times SU(2)$ global symmetry in the bulk,
 $\hat{T}=0$, up to negligible UV-boundary effects. We hence ignore from here on 
 this precision parameter.

\begin{figure}[t]
\begin{center}
\subfigure{\includegraphics[scale=0.5]{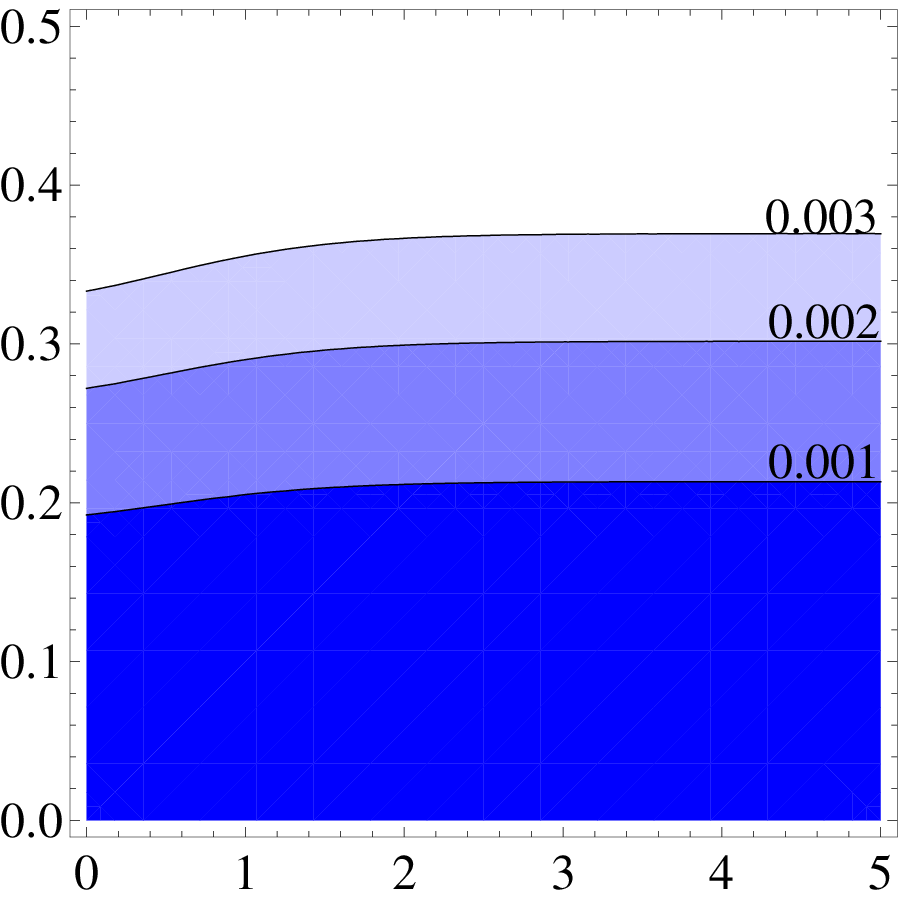}
\Text(-136,80)[]{\small{$\varepsilon$}}
\Text(-54,-5)[]{\small{$r_*$}}}
\subfigure{\includegraphics[scale=0.5]{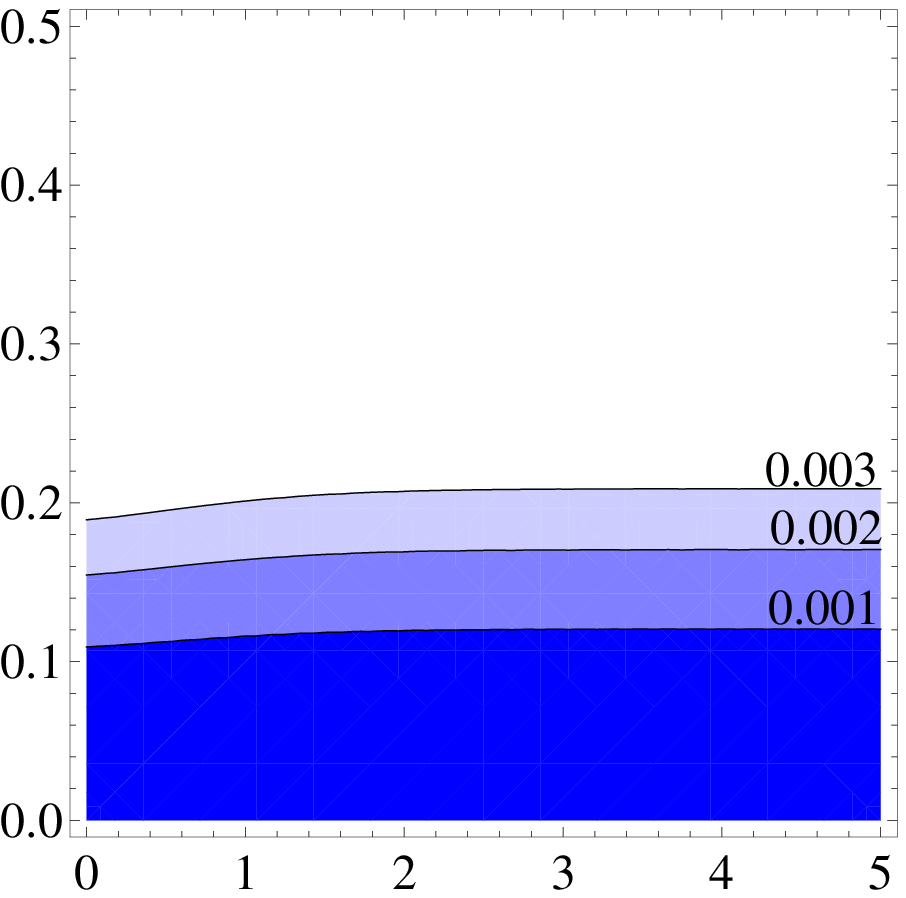}
\Text(-134,80)[]{\small{$\varepsilon$}}
\Text(-54,-5)[]{\small{$r_*$}}}
\subfigure{\includegraphics[scale=0.5]{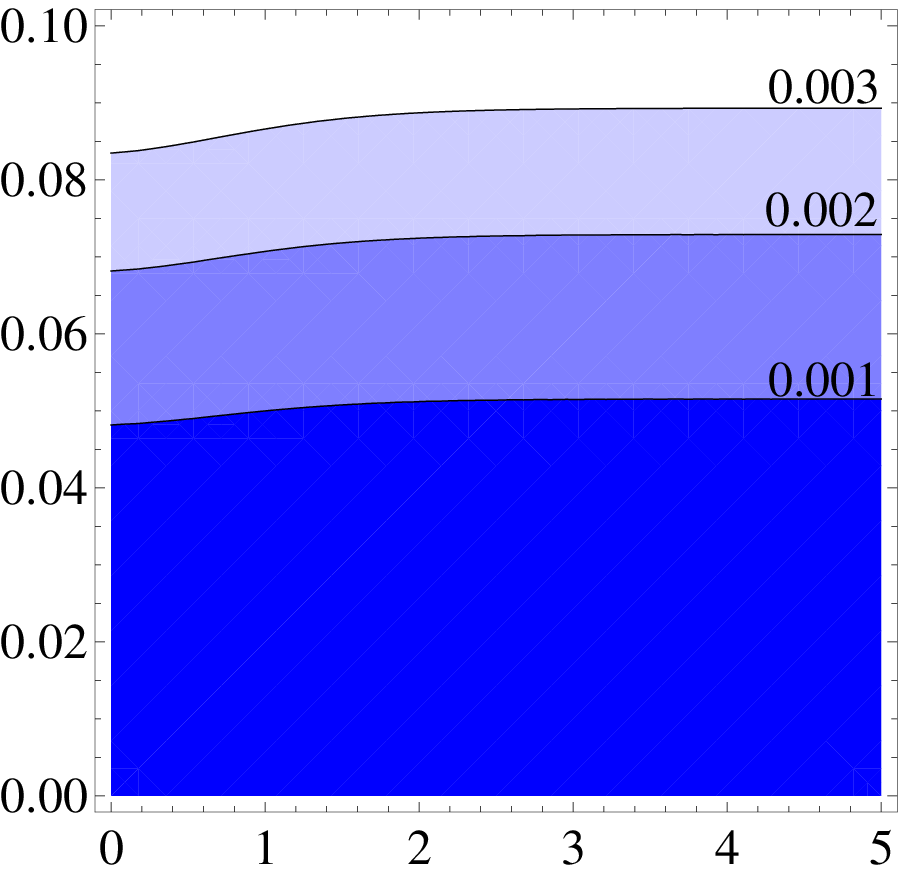}
\Text(-136,80)[]{\small{$\varepsilon$}}
\Text(-54,-5)[]{\small{$r_*$}}}
\subfigure{\includegraphics[scale=0.5]{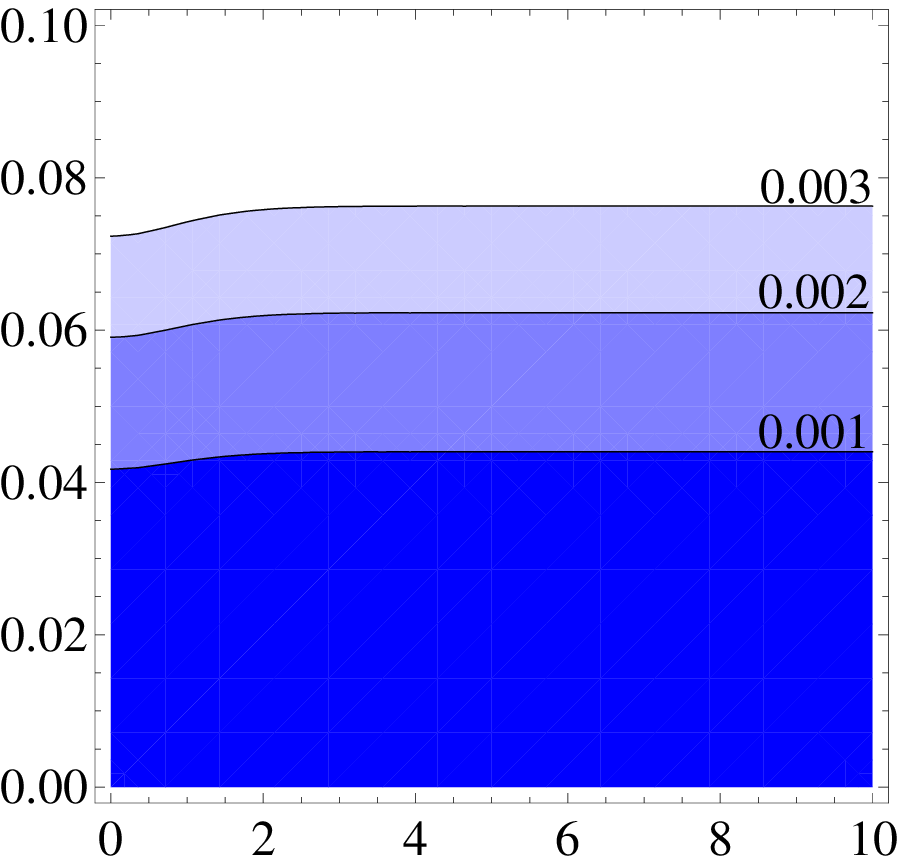}
\Text(-134,80)[]{\small{$\varepsilon$}}
\Text(-54,-5)[]{\small{$r_*$}}}
\end{center}\caption{Plots of $\hat{S}$  against $\varepsilon$ and $r_*$ for $\Omega=0.27,0.5,2,$ and $\Omega=10$ respectively. Note that changing $\Omega$ has a dramatic effect on the bounds on $\hat{S}$. Note also that the scale (in $\varepsilon$) in the last two panels is different from the previous two panels.}
\label{S1}
\end{figure}

\begin{figure}[t]
\begin{center}
\subfigure{\includegraphics[scale=0.5]{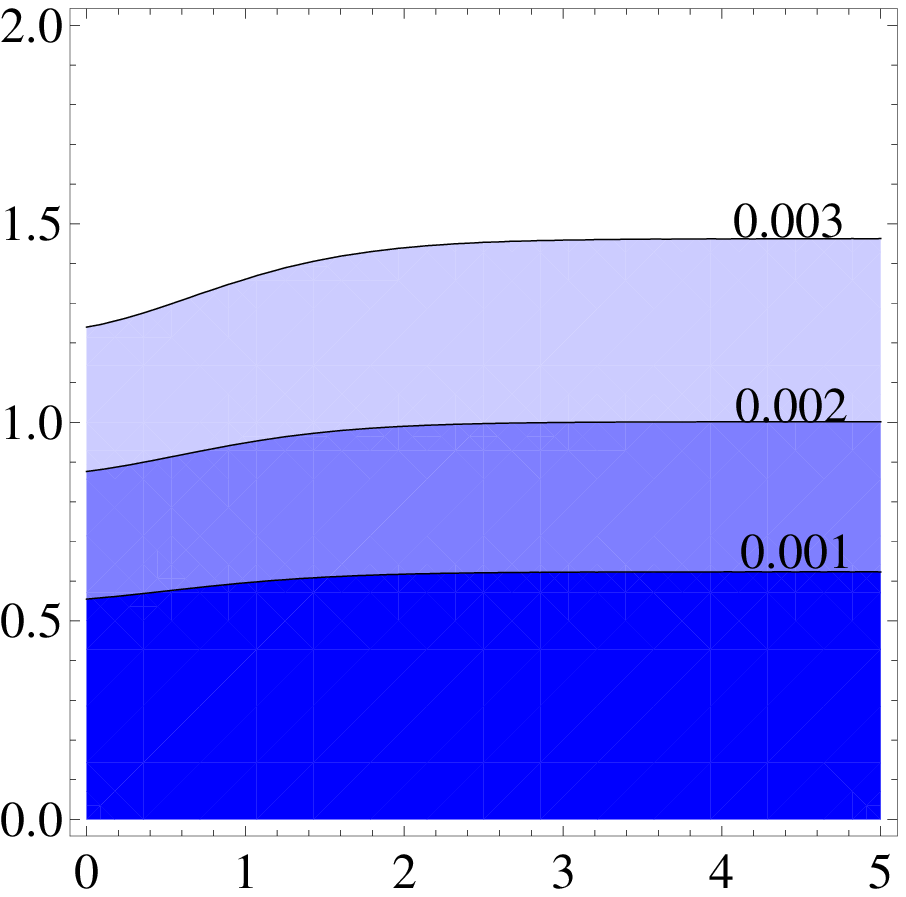}
\Text(-134,90)[]{\footnotesize{$\Omega$}}
\Text(-54,-5)[]{\footnotesize{$r_*$}}}
\subfigure{\includegraphics[scale=0.5]{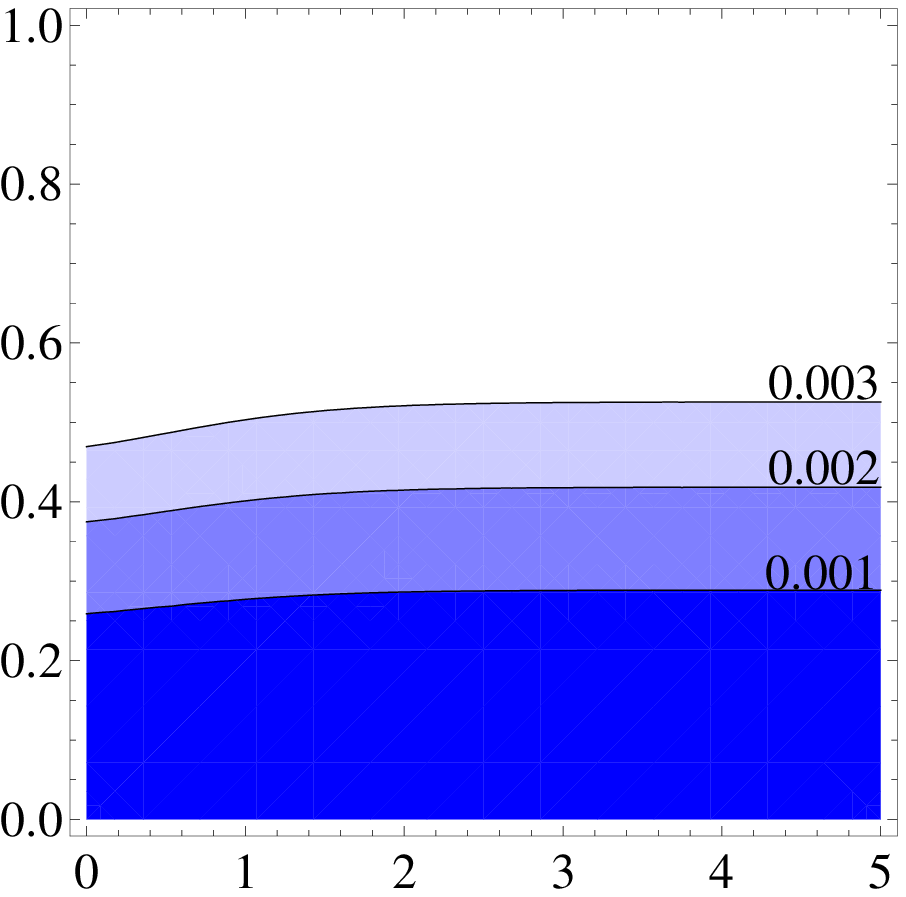}
\Text(-132,90)[]{\footnotesize{$\Omega$}}
\Text(-54,-5)[]{\small{$r_*$}}}
\subfigure{\includegraphics[scale=0.5]{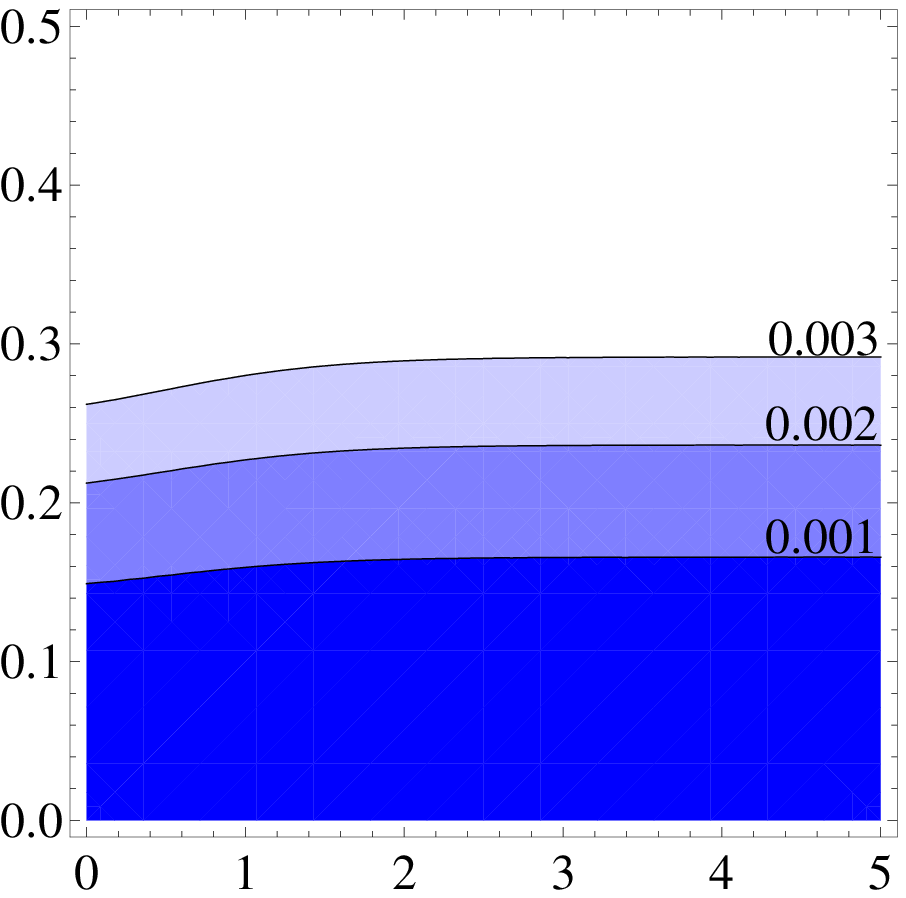}
\Text(-134,90)[]{\footnotesize{$\Omega$}}
\Text(-54,-5)[]{\small{$r_*$}}}
\subfigure{\includegraphics[scale=0.5]{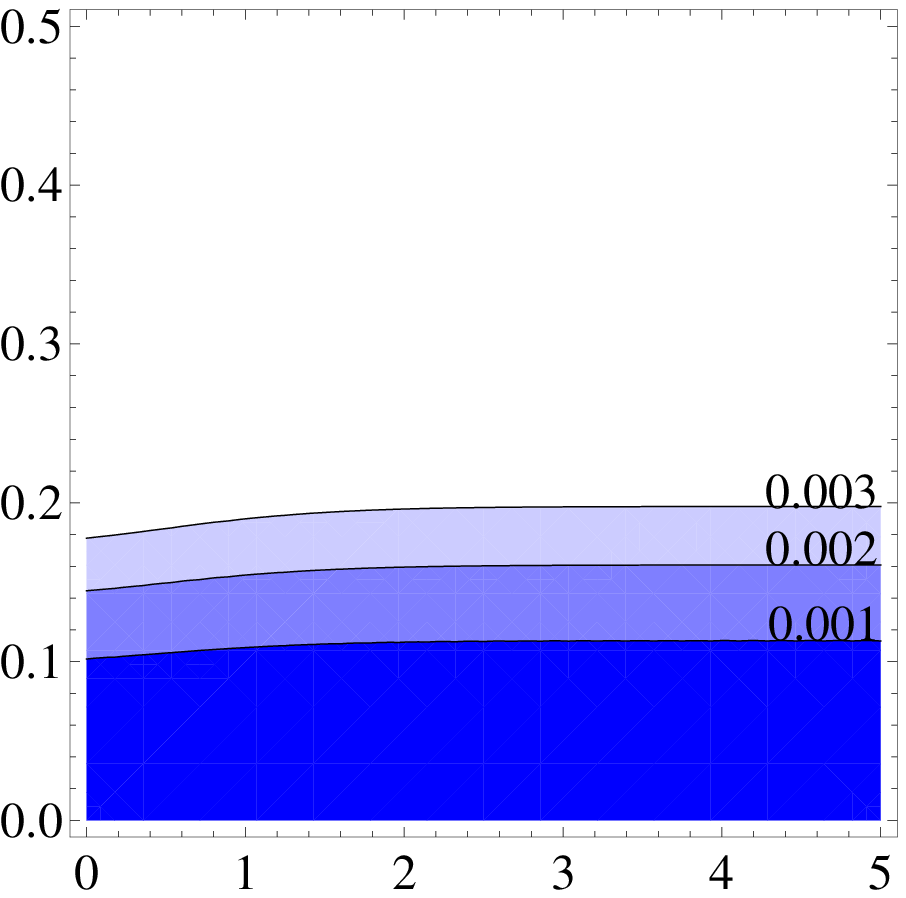}
\Text(-132,90)[]{\footnotesize{$\Omega$}}
\Text(-54,-5)[]{\small{$r_*$}}}
\end{center}\caption{Plots of $\hat{S}$  against $\Omega$ and $r_*$ for $\varepsilon=0.1,0.2,0.343,$ and $\varepsilon=0.5$ respectively. Again note the change in scale. Note also that increasing $\varepsilon$ shifts the allowed values of $\Omega$ down.}
\label{S2}
\end{figure}

\begin{figure}[t]
\begin{center}
\includegraphics[scale=1]{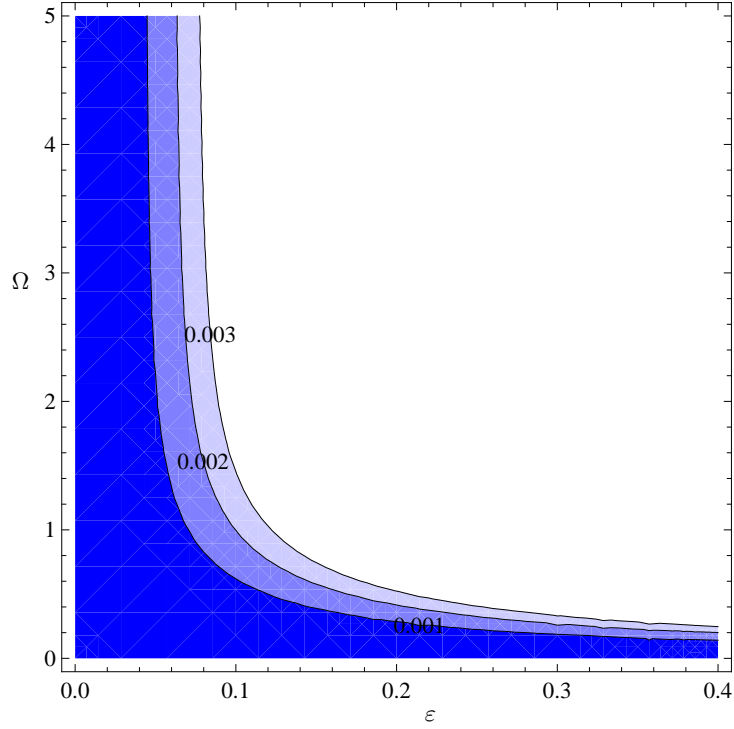}
\Text(-270,160)[]{\small{$\Omega$}}
\Text(-115,-5)[]{\small{$\varepsilon$}}
\end{center}\caption{Plot of $\hat{S}$  
against $\Omega$ and $\varepsilon$ for $r_*=2.5$, $\Delta=1=\Phi_I=e^{r_1}$. 
Note that for $\Omega$ large only small values of $\varepsilon$ 
are allowed by the bounds on $\hat{S}$, while small values of 
$\Omega$ allow for largish values of $\varepsilon$. Note also that changing $r_*$ has little effect on this profile.}
\label{S3}
\end{figure}

Since we have normalized the four-dimensional gauge bosons
 such that $\pi^\prime_{V^3}(0)=1$, $\hat{S}$ measures the deviation of $\pi^\prime_{A^3}(0)$ from unity. Expanding $\pi_{A^3}(q^2)$ as
\begin{equation}
\pi_{A^3}(q^2)=\pi_{A^3}(0)+q^2\pi^\prime_{A^3}(0)+...,
\end{equation}
it is possible to find $\pi^\prime_{A^3}(0)$ and hence $\hat{S}$, which is a  function of the parameters $\varepsilon$, $\Omega$ and $r_*$. When $r_*\lsim 0$, $\hat{S}$ is given approximately by
\begin{equation}
\hat{S}\simeq\frac{\cos^2(\theta_W)\varepsilon^2\Omega^2\left(2+\frac{3}{4}\Omega^2\right)}{(2+\Omega^2)^2},
\end{equation}
which, for small $\Omega,$ becomes
\begin{equation}
\hat{S}\simeq\frac{\cos^2(\theta_W)\varepsilon^2\Omega^2}{2},
\end{equation}
and, for large $\Omega$, becomes
\begin{equation}
\hat{S}\simeq\frac{3\cos^2(\theta_W)\varepsilon^2}{4}.
\end{equation}
 Various plots of $\hat{S}$ are shown in Figures~\ref{S1}, \ref{S2} and~\ref{S3},
 where we show contour plots obtained at fixed values of $\hat{S}$,
 as a function of two of the parameters, with the third one fixed.
 It must be noticed that $\hat{S}$ depends only marginally on $r_{\ast}$,
 particularly when $r_{\ast}\gsim 1$.
 The most interesting results are shown by Fig.~\ref{S3}.
The parameter  $\varepsilon$ is bounded to be 
 rather small ($\varepsilon\lsim 0.07$) if  $\Omega$ is very large, which reproduce the case where electroweak symmetry breaking is induced by Dirichlet IR-boundary conditions for the axial-vector mesons,. 
 However, for smaller values of $\Omega$ larger values of $\varepsilon$ are 
 allowed. For example, for $\Omega\simeq 1/2$ one finds $\varepsilon\lsim 0.18$.

Taking the first term of the expansion of $\pi_{A^3}(q^2)$, and noting that
\begin{equation}
M_Z^2\simeq -\pi_{A^3}(0),
\end{equation}
we find that, as long as $\hat{S}$ is very small, 
so that  $\pi^{\prime}_{A^3}(0)\simeq1$, the mass of the $Z$ is given by
\begin{equation}
M_Z^2=-\frac{4 W_1 \Omega ^2 \varepsilon ^2 e^{2 \left(r_* \left(-\frac{2
   W_1}{3}-1\right)+r_*\right)}}{3 \left(\Omega ^2 \left(e^{-\frac{4 r_*
   W_1}{3}}-\frac{2 W_1}{3}-1\right)-\frac{4}{3} W_1 e^{-\frac{4 r_*
   W_1}{3}}\right)},
\end{equation}
and expression simplifies to
\begin{equation}\label{mz}
M_Z^2\simeq
 \frac{2 \epsilon ^2 \Omega ^2}{\Omega ^2+2},
\end{equation}
for $r_*\simeq 0$. When $\Omega$ is large this becomes
\begin{equation}
M_Z^2\simeq 2\varepsilon^2,
\end{equation}
while for small $\Omega$ we get
\begin{equation}
M_Z^2\simeq \varepsilon^2\Omega^2.
\end{equation}

\section{Mass Spectrum.}

\subsection{Spin-1 states.}

The process of calculating the spin-1 spectrum  consists of finding the 
zeros of the $\pi(q^2)$ functions in Eq.~(\ref{corv}) and Eq.~(\ref{cora}). 
In doing so, one comment is necessary about the spectrum for the charged vector bosons
 (i.e. $(\mathcal{W}_L)_\mu^{1,2}$). 
The SM does indeed have an approximate $SU(2)_L\times SU(2)_R$ symmetry, 
but in the SM only a $U(1)_Y$ subgroup of $SU(2)_R$ is gauged. Also, the mass difference between
top and bottom quarks means that the custodial $SU(2)_R$ symmetry must be broken.
We therefore need to mimic this, which we do by adding 
boundary mass for the $(\mathcal{W}_R)_\mu^{1,2}$ in the UV. 
This change in the UV boundary conditions, shifts the spectrum 
with respect to the (neutral) components in the $T^3$ direction. 
We show this explicitly, by first rotating the $\pi$-functions back to the L-R basis 
and by adding a mass term $m^2$ to the RR component. 
Taking the determinant of this new $2\times 2$ matrix then yields
\begin{equation}
\left|\pi^{LR}_{ij}-\frac{m^2}{2}\left(\begin{array}{cc}0&0\\0&1\end{array}\right)\right|=-\frac{m^2}{2}(\cos^2\theta_w\pi_A+\sin^2\theta_w\pi_V)+\pi_A\pi_V.
\end{equation}
Taking $m=0$ and computing the zeros then gives the neutral spectrum, 
whilst taking $m\rightarrow\infty$ washes out the second term and 
computing the zeros gives the charged sector.
The net result of doing so is that at the level of the light states 
one finds four vector bosons (corresponding to the SM ones), while 
the heavy states form an infinite set of replicas of six states, corresponding to 
the six-dimensional $SU(2)^2$ group.

\begin{figure}[t]
\begin{center}
\subfigure{\includegraphics[scale=0.7]{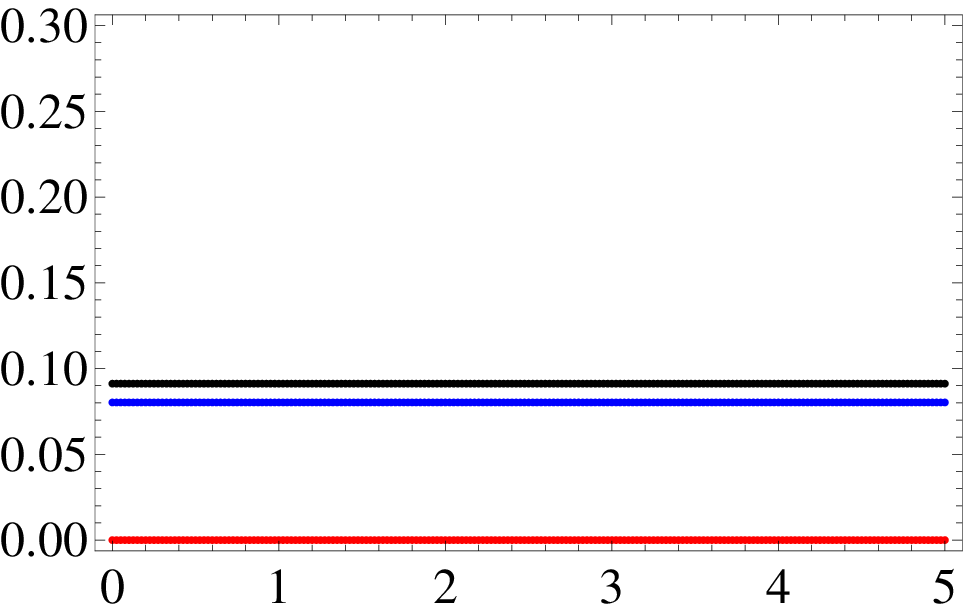}
\Text(-230,107)[l]{\small{$M/TeV$}}
\Text(-15,-5)[]{\small{$r_*$}}}
\subfigure{\includegraphics[scale=0.67]{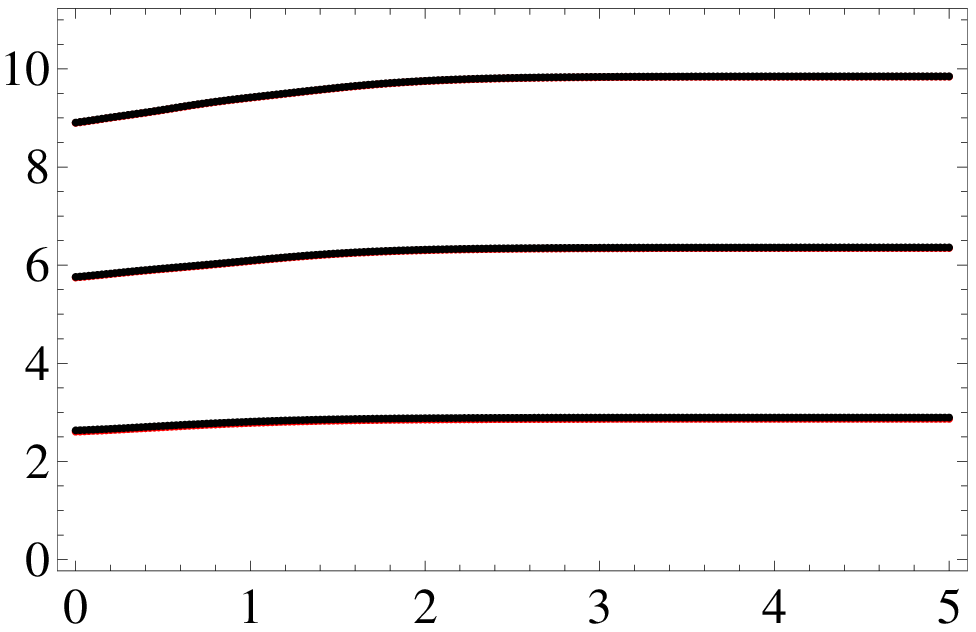}
\Text(10,107)[l]{\small{$M/TeV$}}
\Text(-15,-5)[]{\small{$r_*$}}}
\end{center}\caption{The left panel shows the masses of the light spin-1 states as a function of $r_*$ (vector,axial-vector and charged; i.e. the photon and Z and W bosons). The right panel shows the heavy states. 
We use the choice $\varepsilon=0.34$ and $\Omega=0.27$.
Note that the plot of heavy states includes vector, axial-vector and charged states. 
However, these are all heavily degenerate.}
\label{LS1HS1}
\end{figure}

We first turn our attention to the light states in the spectrum. We already know that there is one massless photon and in addition to this we find a light neutral axial-vector state --- the Z boson, and two charged (left-handed) states --- the $W^\pm$. Fixing the mass of the Z to its experimental value, for all values of $r_*$, yields an overall scale factor that can be used to recover physical units. This is defined as
\begin{equation}
\Lambda_{0}\equiv\frac{M_Z^{exp}}{M_Z}.
\end{equation}
The left panel of Figure \ref{LS1HS1} shows the masses of these light states as a function of $r_*$,
which as anticipated consists of a massless photon, the massive neutral $Z$ and the charged $W$.

We follow the same procedure to calculate the spectrum of heavy states. 
The results are shown in the right  panel of Figure~\ref{LS1HS1}. 
Note that the spectrum of vector, axial-vector and charged states has been plotted. 
However, the towers of  states are quasi-degenerate, 
with only a very small splitting between the masses,
due to the fact that we used a comparatively small value for $\Omega$.
 Also, for small $\Omega$ vector and axial-vector states also become degenerate. 
 The case considered in~\cite{EPdilaton}, 
 which corresponds to $\Omega\rightarrow\infty$, can be thought of as a maximal case of 
 chiral symmetry breaking and produces a spectrum in which the masses of vector and axial-vector states
are very well separated. 
The lightest of the heavy states are the techni-$\rho$ (vector) and techni-$a_1$ (axial-vector), 
each with a mass around $2.3 - 3$ TeV, in agreement with the literature~\cite{AdSTC}. 

\begin{figure}[h]
\begin{center}
\begin{picture}(200,120)
\put(40,5){\includegraphics[height=4.0cm]{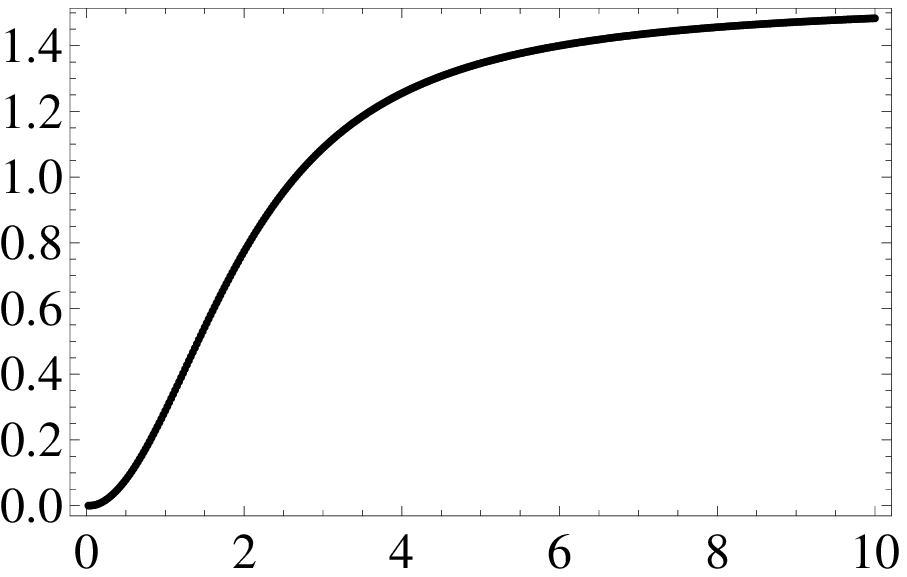}}
\put(0,90){$\frac{M_{A}^2-M_V^2}{M_V^2}$}
\put(190,0){$\Omega$}
\end{picture} 
\caption{Dependence of the mass splitting $(M_{A}^2-M_V^2)/M_V^2$
on the symmetry-breaking parameter $\Omega$, obtained numerically for
$\Phi_I=0$, and $\varepsilon$ tuned so that $\hat{S}=0.003$ for each choice of $\Omega$.}
\label{Fig:massplitting}
\end{center}
\end{figure}

We show in Fig.~\ref{Fig:massplitting} the splitting between the vector and axial vector masses as a 
function of $\Omega$. We remind the reader that in order to avoid the bounds on 
$\hat{S}$, we either have to choose $\varepsilon$ very small, or suppress somewhat $\Omega$.
Notice that, since in the latter case the two states are 
almost degenerate, they cannot be resolved directly in experimental searches.
However, we should expect interference 
between the two states. Were the LHC to observe such states, this interference 
would be seen as a forward-backward asymmetry in the cross-section~\cite{PR}.

\subsection{Scalar Spectrum}\label{secsspec}

To calculate the spectrum of scalars present in this model, one needs 
to consider fluctuations about the classical background. 
Mixing occurs between the fluctuations of $\Phi$ and the scalar fluctuations of the (five-dimensional) 
metric. These fluctuations contain both physical and unphysical degrees of freedom. 
Thankfully, a simple algorithm exists for calculating the 
equations of motion~\cite{BHM} for the gauge-invariant, physical combinations of such fluctuations. 
Put simply, one has to construct gauge invariant variables using the full set of fluctuations, 
expand the equations of motion and boundary conditions
and then replace the physical degrees of freedom with the 
gauge invariant variables. Then finally, the unphysical degrees of freedom may be dropped. 
Applying this process eventually yields the gauge-invariant equations of motion~\cite{BHM}
\begin{equation}\label{gieom}
\left(\left(\partial_r+N-\frac{8}{3}W\right)\left(\partial_r-N\right)+e^{-2A}q^2\right)\mathfrak{a}=0,
\end{equation}
and boundary conditions~\cite{EP}
\begin{equation}\label{eqn:nbc}
\left.\frac{}{}\left(\frac{e^{2A}}{q^2}\frac{(W_\phi)^2}{W}\right)(\partial_r-N)\mathfrak{a}\right|_{r_i}=\left.\frac{}{}\mathfrak{a}\right|_{r_i},
\end{equation}
where $N=W_{\Phi\Phi}-\frac{(W_\Phi)^2}{W}$ and $\mathfrak{a}$ is the gauge-invariant variable, defined by
\begin{equation}
\mathfrak{a}=\varphi+\frac{W_\Phi}{4W}h\,.
\end{equation}
$\varphi$ is the fluctuation of the scalars and $h$ is one of the fluctuations of the metric
(see~\cite{BHM,EP} for details).

\begin{figure}
\begin{center}
\subfigure{\includegraphics[scale=0.7]{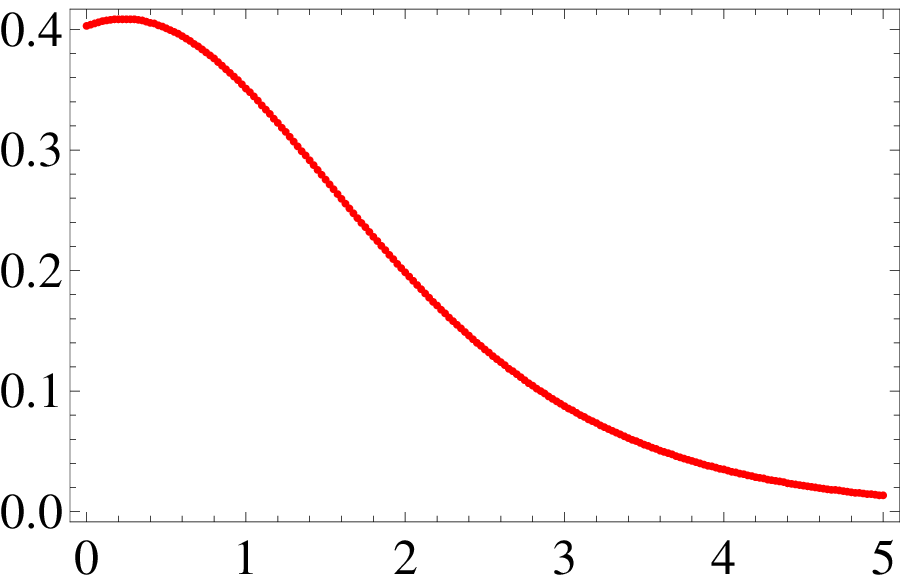}
\Text(-210,102)[l]{\small{$m_d/TeV$}}
\Text(-15,-5)[]{\small{$r_*$}}}
\subfigure{\includegraphics[scale=0.69]{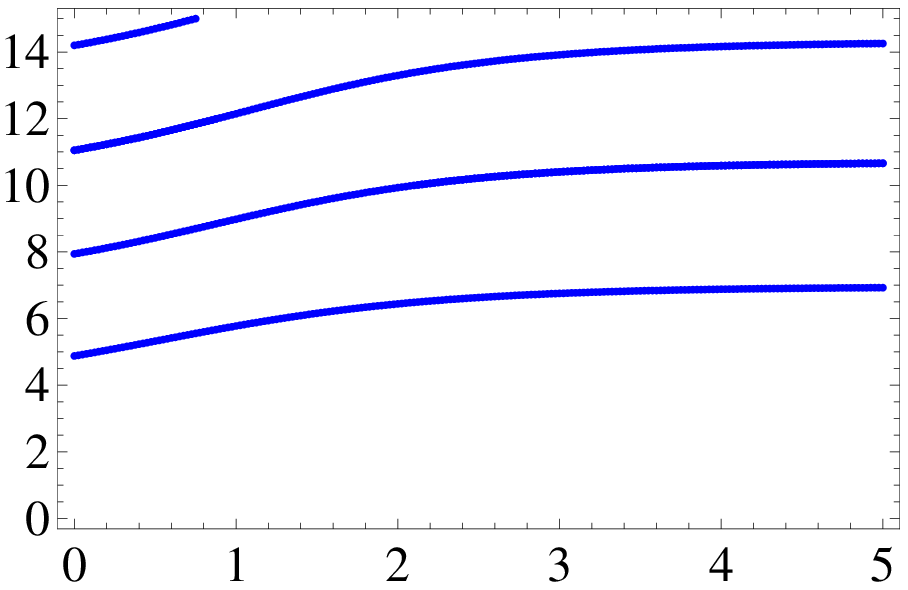}
\Text(10,102)[l]{\small{$M/TeV$}}
\Text(-15,-5)[]{\small{$r_*$}}}
\end{center}\caption{Spectrum of scalar states. 
The left panel shows a light state which is interpreted as a dilaton, 
while the right panel shows the heavy states. Plot obtained by solving numerically 
the equations for $\mathfrak{a}$ with the choice of parameters $\Delta=1=\Phi_I=e^{r_1}$.}
\label{scspec}
\end{figure}

We  use the exact form of the warp factor and solve Eq.~(\ref{gieom}) numerically. 
This yields the spectrum shown in Fig.~\ref{scspec}, which consists of a light state 
which we interpret as a dilaton (left panel) and a tower of heavy resonances (right panel).

\section{Dilaton mass and dilaton-photon-photon coupling.}

The scalar spectrum computed in~\ref{secsspec} contains a light state which 
is to be interpreted as a techni-dilaton. 
This prediction is consistent with the signal, in the LHC data~\cite{LHC}, 
of a light scalar with a mass of $125$ GeV 
and setting the mass of the dilaton to this value fixes $r_*\simeq 2.5$ for $\Delta=1=\Phi_I$.

A very important quantity for LHC searches is the coupling of the light dilaton to 
photons. As in the Standard Model  (and for the same reason) this coupling arises because of 
quantum effects which explicitly break scale invariance.
In contrast to the Standard Model, though, this effects arises at the tree-level in the five-dimensional picture.
In this section we compute explicitly this coupling, and compare it to the 
standard-model coupling of the Higgs particle to two photons.

The starting points are the following terms in the five-dimensional action, which we rewrite here 
explicitly for convenience:
\beqs
{\cal S}&=&
\label{Eq:long}
\int \di^5 x\left\{ \, \sqrt{-g} g^{MN}g^{RS}\left[-\frac{1}{2}\Tr F_{MR}F_{NS}\right]\right.\\
&&\left.\nonumber
-D b(r_2) \delta(r-r_2) \sqrt{-\tilde{g}}\tilde{g}^{\mu\nu}\tilde{g}^{\rho\sigma}
\left[-\frac{1}{2}\Tr F_{\mu\rho}F_{\nu\sigma}\right]
-\sqrt{-g}\frac{1}{2}g^{MN}\partial_{M}\Phi\partial_{N}\Phi\right\}\,.
\eeqs
Here $\tilde{g}$ is the four-dimensional induced metric. Notice that 
$Db(r_2)=r_2-\frac{1}{\varepsilon^2}$ is just a constant, which  earlier on we set so as to 
holographically renormalize  the two-point functions of the gauge bosons.

The perturbed metric can be written as 
\beqs
g_{MN}&=&\left(\begin{array}{cc}
e^{2A} \eta_{\mu\nu}\left(1+\frac{1}{3}h\right) & 0\cr
0 & (1+\nu)^2\end{array}\right)\,,
\eeqs
where $h$ and $\nu$ are scalar fluctuations. All the vectorial and tensorial 
components which play no role in the present discussion have been neglected, 
and we adopt a gauge in which a third scalar fluctuation has been set to zero 
(see~\cite{BHM,EP} and references therein for details).
With all of this, at the linear-order we have
\beqs
\sqrt{-g} g^{11}g^{11}&=&1+\nu\,.
\label{Eq:nu}
\eeqs

The three (in this case gauge-dependent) 
scalar fluctuations $\varphi$, $h$ and $\nu$ can be recombined into
the three physical gauge-invariant fluctuations $\mathfrak{a}$, $\mathfrak{b}$ 
and $\mathfrak{c}$ via~\cite{BHM,EP}
\beqs
h&\equiv&-6A^{\prime}e^{2A}\Box^{-1}\mathfrak{c}\,,\\
\varphi&\equiv&\mathfrak{a}-\bar{\Phi}^{\prime}e^{2A}\Box^{-1}\mathfrak{c}\,,\\
\nu&\equiv&\mathfrak{b}-e^{2A}\Box^{-1}\left(2A^{\prime}\mathfrak{c}+\partial_r \mathfrak{c}\right)\,.
\eeqs
In this new basis, the equation for $\mathfrak{a}$ is the one discussed earlier on, 
while $\mathfrak{b}$ and $\mathfrak{c}$ satisfy the algebraic equations
\beqs
\mathfrak{c}&=&-\frac{2}{3A^{\prime}}W_{\Phi}\left(\partial_r-N\right)\mathfrak{a}\,,\\
\mathfrak{b}&=&\frac{2\bar{\Phi}^{\prime}}{3A^{\prime}} \mathfrak{a}\,,\\
0&=&\partial_r\mathfrak{c}+4A^{\prime}\mathfrak{c}-e^{-2A}\Box \mathfrak{b}\,.
\eeqs
Notice that one of the four equations satisfied by $\mathfrak{a}$, $\mathfrak{b}$ 
and $\mathfrak{c}$ is redundant, being automatically satisfied.

By making use of all of this we find that
\beqs
\nu&=&-\frac{1}{3}h\,,\\
h&=&4e^{2A}\Box^{-1}W_{\Phi}\left(\partial_{r}-N\right)\mathfrak{a}\,,
\eeqs
which will be crucial in the following.
Notice that the first such relation agree with the results by other collaborations (for example compare 
with~\cite{dilaton5D}).

We also expand the Hilbert-Einstein bulk action at quadratic order, 
and making use of the relation between $h$ and $\nu$
we find that
\beqs
\int\di^5 x\sqrt{-g}\frac{R}{4}\,=\,\int\di^5 x \left\{-\frac{1}{24}e^{2A}\partial_{\mu}h\partial^{\mu}h\right\}\,+\cdots\,\,.
\eeqs
We are now ready to start the calculation we are interested in.

The first step is to compute the normalization of the zero-mode of the neutral massless vector meson (photon).
As we saw, it has a wave-function which is flat in the fifth dimension.
The normalization is then given  the first two terms in Eq.~(\ref{Eq:long}), by integrating in the fifth-dimension.
We find:
\beqs
\frac{1}{{\cal N}}&=&\int_{r_1}^{r_2}\di r \left(1-D b(r_2) \delta(r-r_2)\right)\,=\,\frac{1}{\varepsilon^2}\,,
\eeqs
as we saw earlier on, and having set $r_1=0$

The second step consists of computing, from the same two terms in  Eq.~(\ref{Eq:long}),
the overlap integral controlling the cubic coupling. To do so, we use the expansion in Eq.~(\ref{Eq:nu}) 
and obtain
\beqs
v_{d\gamma\gamma}&=&\int_{r_1}^{r_2}\di r \,\nu(r)\,,\\
&=&-\frac{1}{3}\int_{r_1}^{r_2}\di r \,h(r)\,,\\
&=&-\frac{4}{3q^2}\int_{r_1}^{r_2}\di r\,e^{2A} W_{\Phi} \left(\partial_r - N\right)\mathfrak{a}(r,q^2)\,,
\eeqs
where  $q^2=m_d^2$ sets the dilaton on-shell.

The third step requires to normalize the scalar state.
By looking at the kinetic term for $h$,  we define
\beqs
\frac{1}{N}&\equiv&\int_{r_1}^{r_2}\di r \frac{1}{12}e^{2A}h^2\,=\,
\frac{4}{3q^4}\int_{r_1}^{r_2}\di r \, e^{6A} \left[W_{\Phi}(\partial_r-N)\mathfrak{a}(r,q^2)\right]^2
\eeqs

The fourth step requires to reinstate the correct dimensionality.  
Which we do by introducing the scale $\Lambda_0$.
Putting all of this together, we find that the contribution of the strongly-coupled sector to
the coupling between light dilaton $d$ and massless photons reads
\beqs
{\cal L}&=&\frac{1}{2}g_{d\gamma\gamma}\,d\,\Tr F_{\mu\nu}F^{\mu\nu}\,,
\eeqs
with the dimensionful coupling given by
\beqs
g_{d\gamma\gamma}&=&-\frac{v_{d\gamma\gamma} \,{\cal N}\,\sqrt{N}}{\Lambda_{0}}
\,.
\eeqs
What is left is just to perform the numerical integrations yielding $N$ and $v_{d\gamma\gamma}$.
The result can be thought of as the contribution to the low-energy EFT due to loops of 
electrically-charged techni-quarks, which decouple at the strong-coupling scale $\Lambda_{0}\gg m_d$.

However, in order to compare to the data, one has to include also the standard-model contribution,
which is due to (weakly-coupled) one loop diagrams involving the top quark and the $W$ boson.
In order to do so, we need two things: first of all, we need to use the relation between the 
contribution of the top to the Higgs coupling with two photons and to the QED beta-function,
and compare it to the analog results from the strongly-coupled sector.
But also, we need to rescale the results, by keeping into account the fact that the coupling of the SM Higgs
particle is controlled by the electro-weak scale $v_W$, 
while the dilaton is coupled via the  constant $f$.

Let us start from the case in which we have a set of $n_f$ heavy charged fermions with charge $Q$ 
in 4 dimensions coupled to a weakly gauged $U(1)$.
In this case we know that the coupling $\alpha=g^2/(4\pi)$ runs as
\beqs
\alpha(\mu)&=&\frac{\alpha}{1-\frac{\alpha}{3\pi}n_fQ^2 \ln(\mu^2/M^2)}\,,\nonumber
\eeqs
with $\mu$ the renormalization scale and $M$ a reference scale.
The three-level calculation we did for the $U(1)$ in the five-dimensional case yields
\beqs
\alpha(\mu)&=&\frac{\alpha}{1-\frac{\varepsilon^2}{2} \ln(\mu^2)}\,,
\eeqs
which implies the replacement 
\beqs
\beta\,\equiv\,\frac{2\alpha}{3\pi}n_fQ^2&\rightarrow&\beta_5\,\equiv\,{\varepsilon^2}{}\,.
\eeqs

The contribution to the Higgs to two photon coupling in the Standard Model
due to such heavy fermions (the top for instance) can be written as
\beqs
{\cal L}&=& \frac{\beta}{4v_W} h F_{\mu\nu}F^{\mu\nu}\,,
\eeqs
where $v_W$ is the electroweak VEV, which means that we can define the coupling as
\beqs
g_{h\gamma\gamma}&=&\frac{\beta}{v_W}\,,
\eeqs
which means that we can {\it define} the constant  $f$ by the relation
\beqs
g_{d\gamma\gamma}&\equiv&
\frac{\varepsilon^2}{4f}\,.
\eeqs

\begin{figure}[h]
\begin{center}
\begin{picture}(500,420)
\put(10,5){\includegraphics[height=4.5cm]{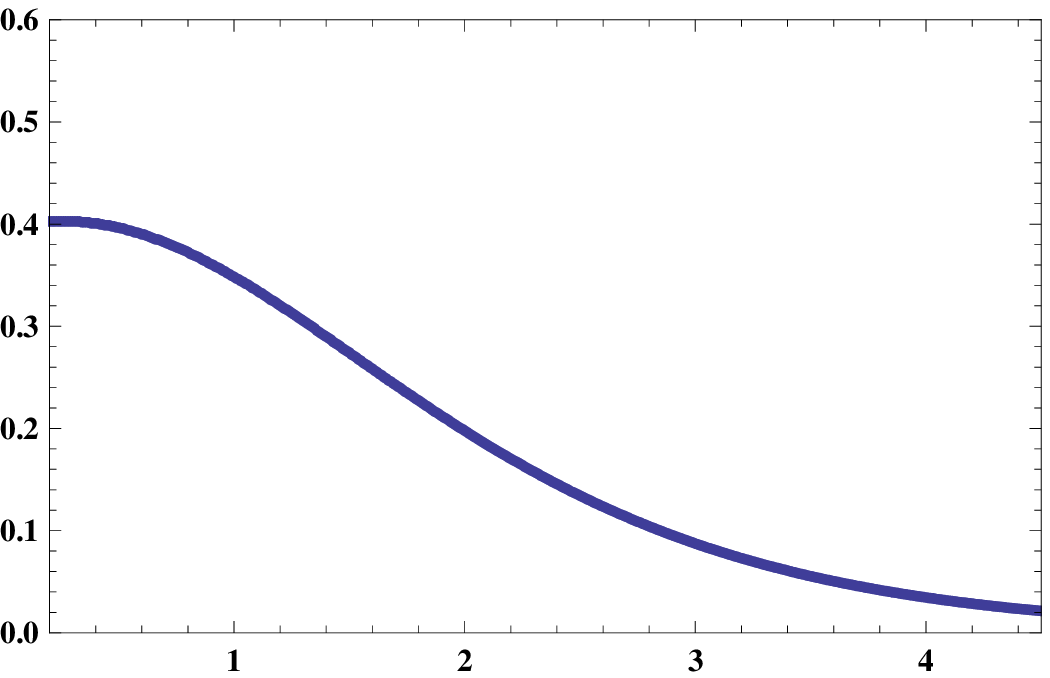}}
\put(10,145){\includegraphics[height=4.5cm]{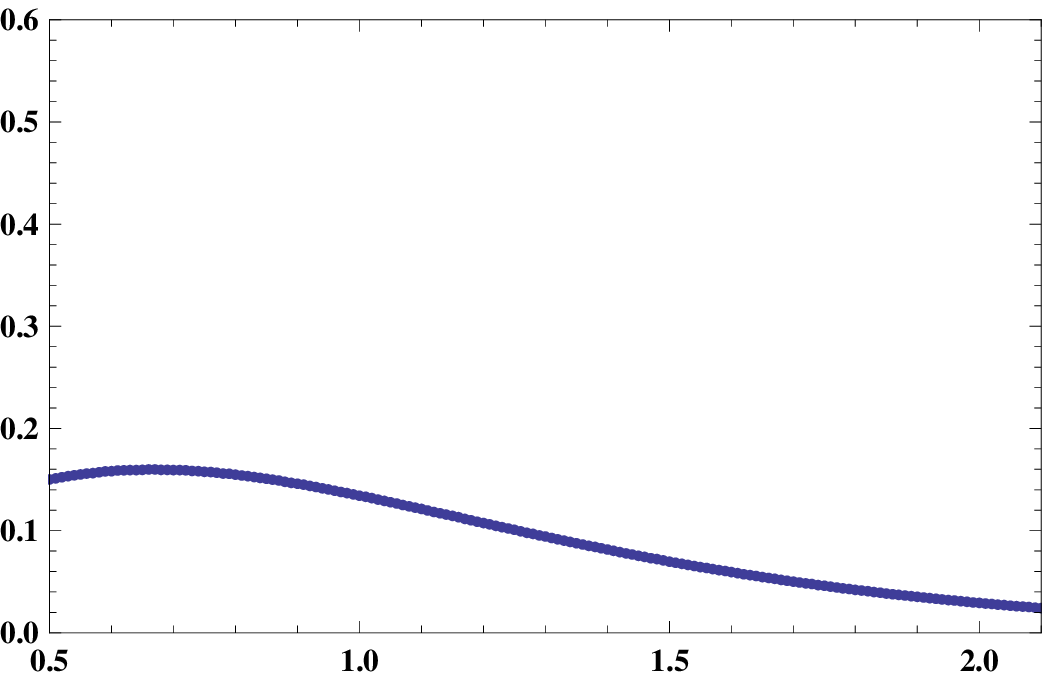}}
\put(10,285){\includegraphics[height=4.5cm]{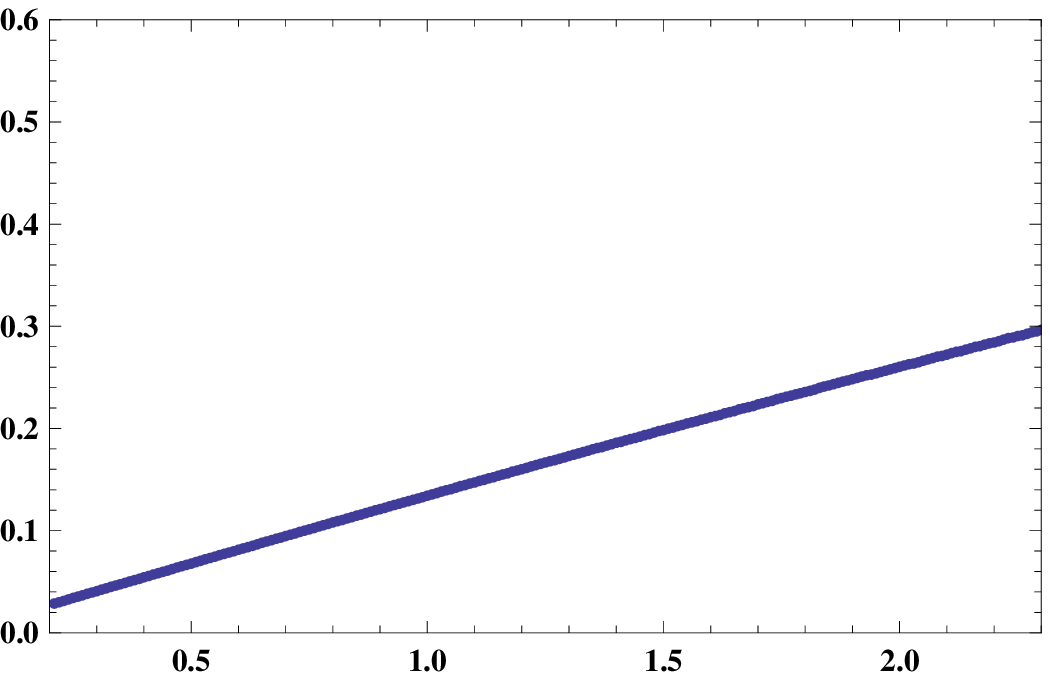}}
\put(260,5){\includegraphics[height=4.5cm]{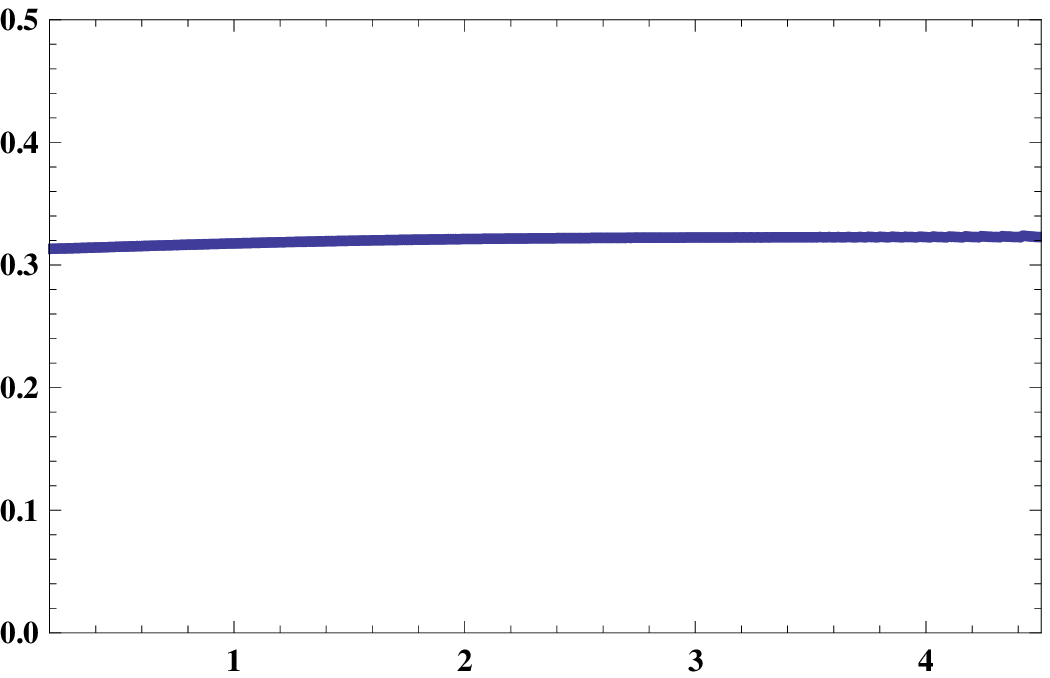}}
\put(260,145){\includegraphics[height=4.5cm]{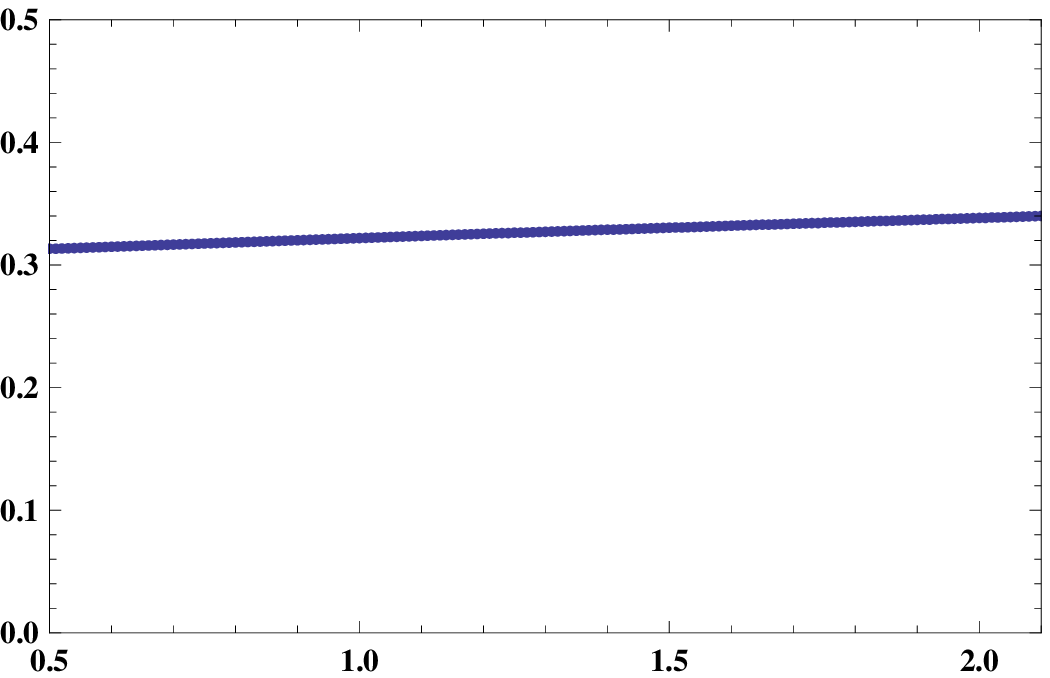}}
\put(260,285){\includegraphics[height=4.5cm]{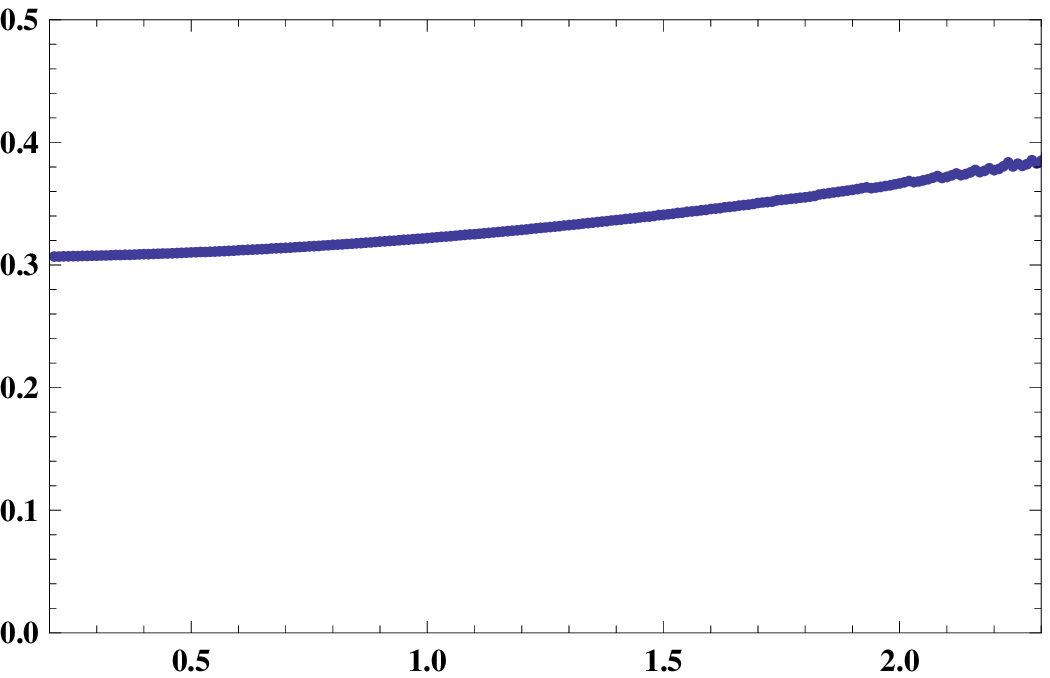}}
\put(0,120){$\frac{m_d}{\Lambda_0}$}
\put(0,260){$\frac{m_d}{\Lambda_0}$}
\put(0,400){$\frac{m_d}{\Lambda_0}$}
\put(250,120){$\frac{f}{\Lambda_0}$}
\put(250,260){$\frac{f}{\Lambda_0}$}
\put(250,400){$\frac{f}{\Lambda_0}$}
\put(200,0){$r_{\ast}$}
\put(200,140){$\Delta$}
\put(200,280){$\Phi_I$}
\put(450,0){$r_{\ast}$}
\put(450,140){$\Delta$}
\put(450,280){$\Phi_I$}
\end{picture} 
\caption{Numerical study of the mass $m_d$ and  constant $f$ (in units of $\Lambda_0$)
of the light pseudo-dilaton. The plots are obtained  by fixing two of the relevant parameters
 to reference values (chosen to be $\Phi_I=1$, $\Delta=1$ and $r_{\ast}=2.5$) and then varying 
 the third. All the plots have been obtained for $r_1=0$ and $r_2=20$.}
\label{Fig:study}
\end{center}
\end{figure}

We need to understand the dependence of the mass and coupling of the dilaton 
upon the parameters $\Delta$, $\Phi_I$ and $r_{\ast}$ before proceeding.
We do so numerically, and the results are reported in Fig.~\ref{Fig:study}.
In the plots, we show the result for the mass $m_d$ and for the  constant $f$,
in units of $\Lambda_0$. The plots show the dependence on $r_{\ast}$  (for $\Delta=1=\Phi_I$),
the dependence on $\Delta$ (for $r_{\ast}=2.5$ and $\Phi_I=1$) and the dependence 
on $\Phi_I$ (for $r_{\ast}=2.5$ and $\Delta=1$).

A few important comments on the figure and on its meaning.
The mass of the dilaton depends (as the mass of the Higgs particle) on the fundamental 
scale characterizing the spontaneous symmetry breaking, and on the size of explicit 
symmetry breaking. The former is determined exclusively by the fact that 
the IR cutoff $r_1$ is always kept fixed to $r_1=0$. We hence expect this scale to be ${\cal O}(1)$
in units of $\Lambda_0$. The latter depends on $\Delta$, $\Phi_I$ and $r_{\ast}$.
In particular, there would be no explicit breaking of scale invariance when 
any of the following is true: $r_{\ast}\rightarrow \pm \infty$, $\Phi_I\rightarrow 0$, $\Delta\rightarrow 0$.
Indeed, the dependence of the mass $m_d$ on these three parameters confirms these statements.

The decay constant should be independent of the explicit symmetry breaking.
However, notice that the way in which we defined it is such that we implicitly reabsorbed the effect
of mixing terms into $f$, which is hence $1/f=\cos\alpha/f_0$, with $f_0$ the true fundamental 
 constant, and $\alpha$ a mixing angle which originates from the fact that the light state
is actually a linear combination of the true dilaton  and of the fluctuations of the 
background scalar. Such mixing is expected to vanish in the same limits in which the dilaton 
is massless, namely when any of the following takes place:  
$r_{\ast}\rightarrow \pm \infty$, $\Phi_I\rightarrow 0$, $\Delta\rightarrow 0$.
Again, in all these limits, $f$ tends to a universal value.
This picture is confirmed by the fact that $f$ grows monotonically away from these limits.
Interestingly, by varying over a reasonable range the three parameters, we find that the change in $f$
is very modest, confirming that the mixing angles are always very small.
We conclude that, in units of $\Lambda_0$ and for $\Phi$ and  $\Delta$ small, but $r_{\ast}$ large, we have
\beqs
\frac{f}{\Lambda_0}&\simeq& 0.3\,.
\eeqs

The reason for this universal behavior is that both the integrals affecting $N$ and $v_{d\gamma\gamma}$
are dominated by the deep IR, where $h\simeq e^{-2r}$.
Hence
\beqs
\frac{f}{\Lambda_0}&=&
-\frac{1}{4v_{d\gamma\gamma}\sqrt{N}}\,\simeq\,
\frac{\sqrt{\frac{1}{12}\int_{r_1}^{r_2}\di r e^{-2r}}}{\frac{4}{3}\int_{r_1}^{r_2}\di r e^{-2r}}\,\simeq\,\sqrt{\frac{3}{32}}\,\simeq \,0.3\,.
\eeqs

At this point, one finds that the decay rate is
\beqs
\Gamma(d\rightarrow \gamma\gamma)&=&\left.\frac{v_W^2}{16f^2}\Gamma(h\rightarrow \gamma\gamma)_{SM}\right|_{\beta\rightarrow \beta+\beta_5}\,
\eeqs
and hence there is a suppression factor due to the fact that the  constant 
$f$ is in general somewhat larger than the electroweak VEV.
The modification to the contribution of the fermions means that there is a further suppression if $\beta_5$
is small (due to the cancellation with the contribution of the $W$), but an 
enhancement for large values of $\beta_5$. 
The former is the expectation of a model where the new-physics sector contains a small number of 
new charged matter fields, while the latter is the generic expectation for a
technicolor theory with $SU(N_T)$ gauge group, in the  large-$N_T$ regime.
Approximately then
\beqs
\frac{\Gamma(d\rightarrow \gamma\gamma)}{\Gamma(h\rightarrow \gamma\gamma)_{SM}}&=&
\frac{v_{W}^2}{16f^2}\frac{\left(\frac{16}{9}-8.3+\frac{4\pi}{\alpha}\frac{\varepsilon^2}{2}\right)^2}{\left(\frac{16}{9}-8.3\right)^2}
\,\simeq\,\frac{v_{W}^2}{16f^2}\frac{(-6.4+800\varepsilon^2)^2}{42}\,,
\eeqs
where we used $\alpha\simeq 1/128$.
As a curiosity, notice that for $\varepsilon\simeq 0.09$, the rate vanishes exactly, irrespectively 
of $f$, because of the cancellation with the $W$ loops.

It is important to notice that the factor of $v_W/4f$ is completely universal, and appears in all 
the amplitudes involving the dilaton. In particular, this factor 
cancels if one takes the ratio of any two decay rates.
The $d\rightarrow \gamma\gamma$, as well as the analog $d\rightarrow gg$,
decay are the only ones the branching ratio of which
is affected by the strong-coupling sector, via the modification of the beta functions
of QCD and QED.

\section{Dilaton-$Z$-$Z$ coupling.}

The coupling to the $Z$ boson differs substantially from the coupling to the photon.
The reason being that there are two  couplings, one of the form
$d F_{\mu\nu}F^{\mu\nu}$ (as for the photon), and one of the form $d A_{\mu}A^{\mu}$
(which exists because of the mass of the $Z$). 
We make use of the fact that expanding at the leading order in the bulk
\beqs
\sqrt{-g}g^{11}g^{55}&=& -e^{2A}\left(1+\frac{1}{3}(h-3\nu)\right)\,=\,-e^{2A}\left(1+\frac{2}{3}h\right)\,,
\eeqs
while for the induced metric $\tilde{g}$ at the boundaries
\beqs
\sqrt{-\tilde{g}}\tilde{g}^{11}&=& -e^{2A}\left(1+\frac{1}{3}h\right)\,.
\eeqs

There are two possible sources of coupling between the scalar fluctuations of the metric and
the massive $Z$ boson. The first, is the same term that appears for the photon. 
We  compute, using the bulk profile $v(r)$ of the $Z$ boson
\beqs
{v}_{dZZ}&=&-\frac{1}{3}\int_{r_1}^{r_2}\di r\, h\, v(r)^2\,,
\eeqs
and the normalization factor
\beqs
\frac{1}{{\cal N}_Z}&=&\int_{r_1}^{r_2}\di r v(r)^2\left(1-D b(r_2) \delta(r-r_2)\right)\,,
\eeqs
and hence obtain the effective coupling
\beqs
g_{dZZ}&=&-\frac{v_{dZZ}{\cal N}_Z \sqrt{N}}{\Lambda_0}\,.
\eeqs
The smallness of the 
 precision parameter $\hat{S}$ implies that ${\cal N}\simeq{\cal N}_Z$,
up to negligibly small corrections $\sim {\cal O}(\hat{S})$.

We need an approximation for $v$ at very small values of the radial direction.
This can be obtained by solving the equation for $\gamma_0$, imposing 
the IR boundary condition $\gamma_0=\Omega^2$ and then integrating to obtain $v$.
We find that 
\beqs
v_0&\simeq&1-\frac{\Omega^2}{2+\Omega^2}e^{-2r}\,,
\eeqs
from which 
\beqs
v_{dZZ}&\simeq&-\frac{1}{3}\int_{r_1}^{r_2}\di r e^{-2r} \left(1-\frac{\Omega^2}{2+\Omega^2}e^{-2r}\right)^2\,\simeq\,c\,v_{d\gamma\gamma}\,,
\eeqs
where $c\rightarrow 1$ for $\Omega\rightarrow 0$ (because in this case the $Z$ and the photon are both massless and obey the same equations), while $c<1$ for generic $\Omega$.
Hence, the result is very similar to the one for the photon,
barring the numerical factor $c$.
In principle, if $\varepsilon\gsim 1$, this might be a very important effect.
In particular, this and the analog coupling to the $W $
 would enhance significantly the vector-boson-fusion
production of the Higgs.
However, this is to be considered just as a curious observation, since 
in a realistic scenario $\varepsilon^2\ll 1$, and hence this coupling only 
amounts to a negligible correction.

The dominant  effect  originates directly from the mass term for the
$Z$ boson, and  is  related to the $h-Z-Z$ coupling of the Standard Model.
It receives two contributions. One from the bulk:
\beqs
\bar{v}_{dZZ}&=&-\int_{r_1}^{r_2}\di r\, \left(e^{2A}\frac{2}{3}h\right)\,2\left(\partial_r v(r)\right)^2\,,
\eeqs
and one coming from the IR-localized mass term from which the IR boundary-conditions arise,
and which yields
\beqs
\tilde{v}_{dZZ}&=&\int\di r \delta(r-r_1)\left(-\frac{1}{3}e^{2A}h v^2\right)\left(2\Omega^2\right)\,.
\eeqs

The resulting effective coupling is then 
\beqs
\bar{g}_{dZZ}&=&-\frac{1}{4}\left(\bar{v}_{dZZ}+\tilde{v}_{dZZ}\right){\cal N}_Z \sqrt{N}\Lambda_0\,.
\eeqs
Notice that the presence of two derivatives in respect to $r$ means that this
has the dimension of a mass. This is indeed the analog of the $h-Z-Z$ tree-level coupling,
which in the Standard Model is $\bar{g}_{hZZ}=M^2_Z/v_W$.

We consider the simplifying case where the AdS curvature in the deep-IR is very close to 1,
so that the space is approximately AdS everywhere. We hence find that the boundary interaction yields
\beqs
\tilde{v}_{dZZ}&\simeq&-\frac{1}{3}v(0)^2\left(2\Omega^2\right)\,.
\eeqs

Hence, by making use of the approximation for $v_0$, we find that
\beqs
\tilde{v}_{dZZ}&\simeq& -\frac{1}{3}\left(\frac{2}{2+\Omega^2}\right)^2\left(2\Omega^2\right).
\eeqs
From the bulk interaction, we obtain (notice that $e^{2A}h\simeq 1$ in the deep-IR, which dominates
the integral)
\beqs
\bar{v}_{dZZ}&\simeq&-\frac{4}{3}\int_{r_1}^{r_2}\di r\, \left(\frac{2\Omega^2}{2+\Omega^2}\right)^2e^{-4r}\,.
\eeqs
Putting the two together:
\beqs
\tilde{v}_{dZZ}+\bar{v}_{dZZ}&=&-\frac{1}{3}\frac{4\Omega^2}{2+\Omega^2}\,.
\eeqs
We can now take into account the normalizations, and find
\beqs
\bar{g}_{dZZ}&\simeq&\frac{1}{3}\left(\frac{4\Omega^2}{2+\Omega^2}\right)\,\varepsilon^2\,{\sqrt{24}}\,\frac{\Lambda_0}{4}\,=\,\frac{M_Z^2}{4\Lambda_0}\,\sqrt{\frac{32}{3}}\,=\,\frac{M_Z^2}{4f}\,.
\eeqs

Hence, we found what we expected from EFT arguments: the  constant $f$ is universal,
and appears both in couplings of the dilaton that originate directly from the masses of the 
light states, and in the couplings that originate at the loop level.
Notice that, as expected, in respect to the standard model the rate of the dilaton decaying into 
$Z$ bosons is suppressed by the $v_W^2/16f^2$ factor, with the only possible caveat coming
from the fact that we neglect the effect of the $g_{dZZ}$ coupling, because we expect 
$\varepsilon$ to be suppressed.

\section{Production Cross-sections.}

The most important production processes for the dilaton,
as for the Higgs particle,  are  the gluon-gluon-fusion and the vector-boson-fusion.
The couplings that control the two rates are suppressed by the universal $v_{W}/4f$ factor.
But both coupling are also affected by enhancement factors due to loops with internal techniquarks.
We already explained that the loop-correction to the h-W-W coupling
is proportional to $\varepsilon^2$, and  is expected to yield a negligible effect,
because $\varepsilon$ cannot be large in a reasonable model.
Hence, this class of models could not explain an enhancement in 
the vector-boson-fusion production rate.

In this paper, we assumed intrinsically that all the techni-quarks be color-singlets under the QCD $SU(3)_c$
gauge group. If so, the only modification to the dilaton-gluon-gluon coupling, in comparison to the SM case,
is the suppression by $v_{W}/4f$.
Yet, the choice of having only color-singlet techniquarks is not only arbitrary,
but also unrealistic, since in any realistic model one needs to couple the 
techni-quarks to the SM fermions in order to give a mass to the latter after the former condense.
 This usually requires to embed ordinary quarks and techniquarks together into 
large representations of a more fundamental Extended Technicolor theory~\cite{ETC,APS},
and hence some of the techni-quarks must be colored.

In order to compute this coupling, one needs to extend the  gauge sector, to allow for the $SU(3)_c$
gauge fields to propagate in the bulk, which is how we would model the fact that techni-quarks are colored.
In order to reproduce the QCD gauge coupling, one has then to holographically renormalize,
which would yield the same algebra as for the photon, but now we would have to introduce a 
different free parameter $\varepsilon_{s}^2$. The net result would be that we would find 
an enhancement controlled by such $\varepsilon_s^2$ for the $dgg$ coupling
in respect to the $hgg$ coupling of the standard model.
Because there are no massive colored gauge bosons, there is no possibility of a cancellation analogous
to the one taking place in the coupling to photons.
But unfortunately this $\varepsilon_s^2$ is a new  free parameter.
The conclusion is that the $gg\rightarrow d$ production cross-section is 
 suppressed by $v_{w}/4f$, and stronglyenhanced by the unknown $\varepsilon_s$.
Hence, we end up with no prediction for this coupling, and for the event rates at the LHC.
In principle, one might be able to relate $\varepsilon_s$ to the physics of additional 
techni-mesons carrying $SU(3)_c$ color, but we will not further explore this possibility in the present paper.

\section{Discussion.}

In this final section, we want to compare our model to the present experimental status of
LHC searches for Higgs-like scalars.
Before doing so, let us summarize briefly our main findings.

\begin{itemize}

\item The model we are considering is characterized by seven free parameters: $\Delta$,
$\Phi_I$, $r_1$, $\Omega$, $\varepsilon$, $r_{\ast}$ and $\Lambda_0$.
The last one has dimension of a mass, and fixes the scale of the theory, while the other six 
control the explicit breaking of scale invariance in the putative  strongly-coupled dual  
theory ($\Delta$, $\Phi_I$ and $r_{\ast}$),
 confinement ($r_1$),  electroweak symmetry-breaking  ($\Omega$),
and the weak gauging of the electroweak symmetry ($\varepsilon$).

\item We impose a set of constraints on the model, based on direct and indirect 
experimental tests of the Standard Model. We impose that $M_Z=M_Z^{exp}\simeq 91$ GeV, 
$m_d=m_s\simeq 125$ GeV and $\hat{S}=\hat{S}_m\simeq 0.003$. 
Because the overall scale of the theory depends on both $\Lambda_0$ and $r_1$,
we can in full generality set $r_1=0$. 
Because $\Delta$, $\Phi_I$ and $r_{\ast}$ all enter the expression for the mass of the dilaton,
but affect only very modestly all other interesting quantities
we will  fix $\Delta$ and $\Phi_I$ 
to indicative, generic values ${\cal O}(1)$.
As a result, the allowed parameter space is spanned by only one free parameter, which we may
choose to be $\varepsilon$ or (equivalently) $\Omega$.

\item There are four predictions that can be made, after imposing all of these constraints. 
These are the $d\rightarrow \gamma\gamma$ and $d\rightarrow ZZ$ decay rates, 
which can be compared to the standard-model analogs, the overall mass scale  $\Lambda_0$
of the heavy resonances,
 and the mass splitting between the towers of heavy spin-1 resonances,
which depends crucially on $\Omega$.

\item The measurable quantities directly related to electroweak symmetry breaking $M_Z$ and $\hat{S}$
are barely sensible to $r_{\ast}$, as long as $r_{\ast}\gsim 1$ (and $\Delta$ and $\Phi_I$ are ${\cal O}(1)$). 
We can use the approximate relations
\beqs
M_Z^2&\simeq &
\frac{2\varepsilon^2\Omega^2}{\Omega^2+2}\Lambda_0^2\,,\\
\hat{S}&\simeq &\frac{\varepsilon^2\Omega^2(2+3\Omega^2/4)}{
(2+\Omega^2)^2}\cos^2\theta_W\,,
\eeqs
where we neglected the effect of the change of curvature in the deep IR. 

\item The mass $m_d$ of the lightest scalar depends on $\Lambda_0$ and on $r_{\ast}$.
Imposing the constraint that $m_d\simeq 125$ GeV renders $\Lambda_0=\Lambda_0(r_{\ast})$,
and hence this is what determines the overall physical scale of the model.
Because the other approximations are valid provided $r_{\ast}\gsim 1$, we 
restrict our attention to this regime, and to $\Delta=1=\Phi_I$,
 for which the mass of the lightest state is well approximated by:
\beqs
m_d&\simeq&1.9\,\Lambda_0\, e^{- r_{\ast}}\,.
\eeqs

\item As long as we are not too far from the limit in which the lightest state is
an exactly massless dilaton, we can approximate 
\beqs
f&\simeq&0.3\,\Lambda_0\,.
\eeqs

\item The decay rate of the dilaton into two photons compares to the SM
decay of the Higgs particle as
\beqs
\frac{\Gamma(d\rightarrow \gamma\gamma)}{\Gamma(h\rightarrow \gamma\gamma)_{SM}}&\,\simeq &\frac{v_{W}^2}{16f^2}\frac{(-6.4+800\varepsilon^2)^2}{42}\,,
\eeqs
while for the decay to $ZZ$ or $WW$ (including the cae when the gauge bosons are off-shell,
and hence the final state consists of SM fermions)
\beqs
\frac{\Gamma(d\rightarrow ZZ)}{\Gamma(h\rightarrow ZZ)_{SM}}&\,\simeq &\frac{v_{W}^2}{16f^2}\,.
\eeqs
In this we assumed that $\varepsilon$ is small, so as to ignore the 
(techini-quark) loop-induced possible enhancement of the $d\rightarrow ZZ$ rate. This also implies
that we do not expect the vector-boson-fusion production mechanism to be enhanced.

\end{itemize}

With all of the above put in place, we can now provide some examples of the results,
within these approximations.
First of all, we solve for $\Lambda_0$, impose $m_d=m_s$, and find
\beqs
\Lambda_0&\simeq&\,e^{r_{\ast}}\,(66 {\rm GeV})\,.
\eeqs

We can then extract $\varepsilon^2$ from the expression for $\hat{S}$, and impose 
that $\hat{S}$ be at the boundary of the experimentally allowed range (hence maximizing the 
possibility that new physics be detectable):
\beqs
\varepsilon^2&\simeq&\frac{(2+\Omega^2)^2}{\Omega^2(2+3\Omega^2/4)}\,(0.0045)\,.
\eeqs
By replacing in the expression for $M_Z$ we find that
\beqs
\frac{(2+\Omega^2)}{(2+3\Omega^2/4)}&\simeq&200 e^{-2r_{\ast}}\,.
\eeqs

By choosing $\Omega\rightarrow +\infty$
one is forced to choose $r_{\ast}\simeq 2.5$, and hence $\varepsilon^2\simeq 0.006$
and $\Lambda_0\simeq 800$ GeV, in good agreement with 
the exact results obtained numerically in~\cite{EPdilaton}.

On the other extreme, taking $\Omega\ll 1$ affects only modestly the estimate of $r_{\ast}\simeq 2.65$,
yielding $\Lambda_0\simeq 930$ GeV. 
At this point one is left with the conclusion that $r_{\ast}$ and $\Lambda_0$ can vary only very 
moderately after imposing the constraints from $\hat{S}$, $M_Z$ and $m_s$,
but there is freedom in choosing $\Omega$ (and consequently $\varepsilon$).
In particular, $\varepsilon$ is allowed to take values larger than $0.006$.
We should notice that values of $\varepsilon\gsim 1$ 
are excluded on the basis of subleading precision parameters such as $W$ and $Y$.
Also, we must remind the reader that we have fixed $\hat{S}$ to its largest allowed value.

\begin{figure}[h]
\begin{center}
\begin{picture}(400,240)
\put(50,5){\includegraphics[height=7.0cm]{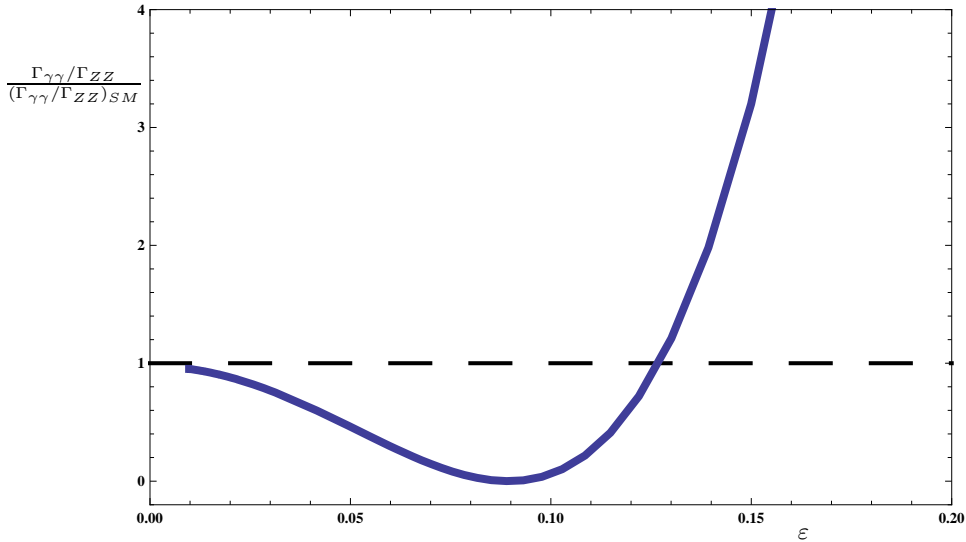}}
\put(0,170){$\frac{\Gamma_{\gamma\gamma}/\Gamma_{ZZ}}{(\Gamma_{\gamma\gamma}/\Gamma_{ZZ})_{SM}}$}
\put(300,0){$\varepsilon$}
\end{picture} 
\caption{Ratio of the decay rates $d\rightarrow \gamma\gamma$  and $d\rightarrow ZZ$
normalized over the SM decay rates of the Higgs particle
onto the same final state, for $m_d=125$ GeV, $\Lambda_0\simeq 900$
 GeV, $\Delta=e^{r_1}=\Phi_I=1$, $r_{\ast}\simeq 2.4-2.6$,
as a function of $\varepsilon$, with $\Omega$ dialed to reproduce the mass of the $Z$ boson $M_Z\simeq 91$ GeV. The black (long-dashed) line is the Standard Model normalized to $1$, 
 the blue (thick) line is this model as a function of $\varepsilon$.}
\label{Fig:rate}
\end{center}
\end{figure}

In Fig.~\ref{Fig:rate}, we show one  example illustrating how the 
decay rate of the dilaton into two photons depends on $\varepsilon$.
Suppression is present for small-to-moderate
 values of $\varepsilon$, due to the cancellation between the new physics contribution and the 
 standard-model one. For values of $\varepsilon\gsim 0.12$, we find an enhancement, 
 which is $\propto \varepsilon^4$. Notice that  the production 
 cross-section in gluon-gluon-fusion is controlled by an unknown parameter $\varepsilon_s$,
 which drops out of the double ratio in the Figure.
 
We conclude this part of the analysis by asking ourselves what is the reasonable range of values 
which is acceptable for $\varepsilon$.
We already reminded the reader of the fact that very large values of $\varepsilon$ would affect 
the subleading precision parameters $W$ and $Y$.
There is a direct way of seeing why taking arbitrarily large values of $\varepsilon$ yields to a problem.
 By looking back at the beta-function of the QED gauge coupling, 
 and neglecting the SM contribution while focusing
 on the new-physics one, we recognize the presence of a pole for 
 \beqs
 \log\left(\frac{\bar{\mu}}{M}\right)=\frac{1}{\varepsilon^2}\,,
 \eeqs
where $M$ is of the order of the electroweak scale.
For the calculations we performed regarding the electroweak precision parameters to be useful,
we need the scale $\bar{\mu}$ to be at least larger than the first few heavy resonances,
otherwise the Landau pole would emerge before that and invalidate our analysis.
In practical terms, $\varepsilon^2 \gsim 1$ is the regime in which the low-energy EFT
is itself strongly-coupled, and hence useless.
By requiring $\bar{\mu}/M\gsim 15$,
we find $\varepsilon\lsim 0.6$.
For $\varepsilon\simeq 0.15$, we find that $\bar{\mu}/M\simeq e^{1/\varepsilon^2}\simeq 10^{19}$
is safely larger than the physics scales of interest here.  The decay into two photons 
is enhanced by a factor of two in respect to the standard model in this latter case.

Let us ask what are the smallest allowed values of $\varepsilon$
by assuming that this is the actual dual of a very simple technicolor model, with
some number $n_f$ of electrically charged techni-quarks.
Because the bulk gauge symmetry is $SU(2)_L\times SU(2)_R$, there is no $U(1)_{B-L}$,
hence the charge of such fermions must be $Q=\pm 1/2$. 
By taking literally the relation
\beqs
\frac{\varepsilon^2}{2}&=&\frac{\alpha}{3\pi}n_f Q^2\,,
\eeqs
we get that
\beqs
n_f&\simeq&2400 \varepsilon^2\,,
\eeqs
which means that for $\varepsilon^2\simeq 0.006$ (which is the bound from $\hat{S}$
in the $\Omega\rightarrow +\infty$ case), then  $n_f=12$.
This translates into a number $N_D=n_f/2=6$ of new doublets of $SU(2)_L$,
which is in reasonable agreement with extrapolations of 
perturbative calculations of the $\hat{S}$ parameter~\cite{Peskin}.
If the dual theory is a $SU(N_T)$ gauge theory, and the techni-fermions are
on the fundamental, this means that this limiting value of $\varepsilon$ would
correspond to $N_T=6$, provided there is only one such $SU(N)_T$-fundamental, 
$SU(2)_L$-doublet Dirac fermion. 
In spite of the many questionable 
identifications we made in this paragraph (all the equations of this paragraph should be taken with a
grain of salt), we learn that if is reasonable to take 
$\varepsilon^2\gsim$ few $10^{-3}$, and that the realistic range 
within which $\varepsilon$ is allowed to vary is $\varepsilon^2\gsim$ few $10^{-3}$
and $\varepsilon^2\lsim$ few $10^{-1}$ .  
The lower bound being given by the comparison with a gauge theory,
the upper bound by the necessity to avoid a Landau pole at very low scales.

Near the lower bound allowed for $\varepsilon$
the dual theory has a very small number of degrees of freedom,
and hence the phenomenology at very low energies resembles that of a weakly-coupled model.
In particular, all the decay rates of the dilaton are  suppressed in respect to the Standard Model.
At the same time, there is no special symmetry pattern emerging in the spectrum of heavy 
composite states, which have masses and mass-differences all of order $\pi \Lambda_0$.
Conversely, at the upper end of the allowed range for $\varepsilon$
 the dual theory is a  large-$N_T$ theory, and correspondingly we expect an enhancement of the 
 $h\rightarrow \gamma\gamma$ decay rate. The bound on the $\hat{S}$ parameter is satisfied only
 because $\Omega\ll 1$, which in turns implies that the heavy spin-1 resonances are almost insensitive
 to electro-weak symmetry breaking, the mass splitting between the lightest vector and axial-vector 
resonance being parametrically suppressed (see Fig.~\ref{Fig:massplitting}).

A final remark is related to the $\hat{S}$ parameter. We obtained the numerical examples illustrated 
above by requiring that $\hat{S}=0.003$ lies at the boundary of what allowed by precision physics.
If one were to tighten this bound, the result would be that the heavy resonances 
become undetectable because they would be too heavy. At the same time,  
this would make also the decay constant $f$
larger, and hence strongly suppress all the couplings 
of the dilaton in respect to those of the SM Higgs.
A strong suppression of all the production rate  and decay rates 
would emerge, and the dilaton would become very narrow.

\section{Conclusions}

Let us briefly summarize the content of the paper.
We considered the bottom-up description of a confining, approximately scale-invariant
strongly-coupled model of electro-weak symmetry breaking,
by constructing a five-dimensional background obtained as 
a classical solution of a five-dimensional scalar theory coupled to gravity.
We chose the dynamics of the scalar in such a way that the geometry 
in the fifth dimension smoothly interpolates between two 
different AdS regions, with different curvature.
Confinement is modeled crudely with an IR cutoff.
The explicit breaking of scale invariance is due to the fact that at some scale $r_{\ast}$ 
the warp factor changes.
The SM gauge group is introduced by allowing $SU(2)_L\times SU(2)_R$ gauge bosons
to propagate in the bulk.
Holographic renormalization allows to weakly gauge the $SU(2)_L\times U(1)_Y$ sub-group,
and the parameter $\varepsilon$ controls the strength of the resulting gauge coupling
in respect to the self-coupling of the spin-1 heavy resonances.

One important difference in the basic action in respect  to~\cite{EPdilaton},
is that here we generalized the way in which EWSB is broken. 
Instead of imposing Dirichlet boundary conditions
for the axial-vector excitations, we allow for generalized 
Neumann boundary conditions, controlled by a symmetry-breaking 
parameter $\Omega$, such that for $\Omega\rightarrow +\infty$ we 
recover the Dirichlet boundary conditions,
while for $\Omega\rightarrow 0$ there is no EWSB.

We computed the precision parameter $\hat{S}$ in this generalized scenario, and the spectrum of spin-1 states.
We also repeated the calculation of the scalar excitations, confirming the results in~\cite{EPdilaton},
in particular the fact that the spectrum naturally contains a parametrically light pseudo-dilaton.
We computed the decay constant $4f$ of such dilaton, and its coupling to two photons and two $Z$ bosons.
We compared the resulting decay rates to the decay rates of the Higgs particle of the minimal version of the Standard Model, as a function of the parameters
of the present model, in particular $\varepsilon$, $r_{\ast}$ and $\Omega$.
We also discussed the effects on the production cross-sections, in particular the
gluon-gluon-fusion and vector-boson-fusion processes.

We find that the vector-boson-fusion production cannot be enhanced, 
while the gluon-gluon-fusion is effectively a free parameter.
For the decay rates, we find a universal suppression factor $(v_W/4f)^2$,
but that the $d\rightarrow \gamma\gamma$ decay rate is also a  non-trivial function of $\varepsilon$.
In particular, by taking the ratio of the number of events with two photons to the number of events with two
(virtual or real) $Z$ or $W$ bosons, all the other parameters drop,  
and only the dependence on $\varepsilon$ is left, which can hence be extracted from the data
unambiguously.

 It is possible to make choices of $\varepsilon$ that allow to reproduce for the 
 dilaton results which are very similar to those of the
  Higgs particle in the minimal version of the Standard Model,
 in particular by choosing $\varepsilon$ to be very small (corresponding to 
 a dual technicolor theory with few new fundamental degrees of freedom), 
 or close to $\varepsilon\simeq 0.12$.
A suppression of the number of events with two photons is also possible, 
provided $\varepsilon\lsim 0.12$, due to the cancellations with the contribution
to the coupling due to loops of $W$ bosons.

However,  for the natural choices of $\varepsilon\gsim 0.12$
(in which case cancellations are not important and the dual theory is in the large-$N$
regime most suited to holographic calculations)  the 
$d\gamma\gamma$ coupling is strongly enhanced.
Hence this model would accommodate very easily an enhancement of the
decay rate of a new particle with $125$ GeV mass into a final state with two photons,
in respect to the other decay rates, in comparison  to the SM predictions for the Higgs particle.
Interestingly, the phenomenology of the spin-1 composite states is very different 
depending on $\varepsilon$, and hence measuring this parameter from the decays of the dilaton
allows to predict the mass differences between vector and axial-vector resonances,
and to test the model.

Note added: on July 4th, 2012, ATLAS and CMS announced $\sim 5\sigma$ evidence
in the data for a boson with mass $m_s\sim 125$ GeV.
The observed event rates are compatible both with the Standard Model 
and with this model, given the current experimental uncertainty.
A more precise measurement of the ratio between number of decays
in two photon and two heavy (virtual) gauge bosons  might allow
to discriminate between the two.

\vspace{1.0cm}
\begin{acknowledgments}
We thank Adi Armoni and Daniel Elander for useful discussions and comments on the manuscript.
The work of MP is supported in part by WIMCS and by the STFC grant ST/J000043/1.
RL is supported by the STFC Doctoral Training Grant ST/I506037/1.
\end{acknowledgments}


\end{document}